%% file: main2.tex
\documentclass[preprint,3p,authoryear]{elsarticle}
\usepackage[utf8]{inputenc}
\usepackage{amsmath,amsfonts,amssymb,amsthm}

\usepackage{color}
\usepackage{eurosym}
\usepackage{mathrsfs}
\usepackage{multirow, microtype}
\usepackage{geometry, array}
\usepackage{tikz}
\usetikzlibrary{shapes.geometric, arrows}
\tikzstyle{Userchoice}=[rectangle, rounded corners,text centered, draw=black]
\tikzstyle{RECchoice}=[rectangle, text centered, draw=black]
\tikzstyle{Randomechoice}=[diamond, aspect=2,text centered, draw=black]
\tikzstyle{STchoice}=[text centered,inner sep=1]
\tikzstyle{arrow} = [thick,->,>=stealth]
\usepackage{pgfplots}
\pgfplotsset{compat=1.18}
\usepackage{url}
\usepackage{colortbl}
\usepackage{mathtools}
\usepackage[breakable]{tcolorbox}
\usepackage[font=normalsize]{caption}
\usepackage[caption=false,font=footnotesize]{subfig}
\usepackage{makecell}
\usepackage{enumitem}

\newtheorem{theorem}{Theorem}
\newtheorem{corollary}{Corollary}

\newcommand{\bb}[1]{\mathbb{#1}}
\newcommand{\set}[3][]{\left\lbrace #2 \ #1\vert \ #3 \right\rbrace}

\newcommand{\users}{\ensuremath{\mathcal{N}}}
\newcommand{\ConstSet}{\ensuremath{\Omega}}
\newcommand{\newusers}{\ensuremath{\mathcal{M}}}
\newcommand{\ppar}{\vspace{1.5ex}}

\usepackage[acronym,nonumberlist,nogroupskip,nowarn]{glossaries}
\makeglossaries
\input{acronyms}

\begin{document}
\begin{frontmatter}
\title{A Decision-Making Framework for New Member Integration in Renewable Energy Communities under Prospect Theory}
\author[label1,label3]{Louise Sadoine}\corref{cor1}
\ead{louise.sadoine@kuleuven.be}
\author[label2]{Thomas Brihaye}
\ead{thomas.brihaye@umons.ac.be}
\author[label1]{Zacharie De Grève}
\ead{zacharie.degreve@umons.ac.be}

\affiliation[label1]{organization={Power Systems and Markets Research Group, University of Mons},
           addressline={Boulevard Dolez 30},
            postcode={7000},
             city={Mons},
             country={Belgium}}

\affiliation[label2]{organization={Effective Mathematics Department, University of Mons},
            addressline={Avenue Victor Maistriau 15},
            postcode={7000},
            city={Mons},
            country={Belgium}}

\affiliation[label3]{organization={Electrical Energy Systems and Applications, KU Leuven and EnergyVille},
            addressline={Kasteelpark Arenberg 10},
            postcode={3001},
            city={Leuven},
            country={Belgium}}

\cortext[cor1]{Corresponding author}

\begin{abstract}
This paper introduces an original approach to an underexplored issue: the integration of a new member into an existing renewable energy community. The problem involves actions with both long-term consequences, such as investment and local pricing, and short-term operational ones, such as daily energy and financial flow management. Long-term decision-making is modeled using finite extensive-form game theory, while short-term day-ahead scheduling decisions are formulated as a generalized Nash equilibrium problem. This framework explicitly accounts for heterogeneous stakeholder preferences and bounded rationality, modeled through prospect theory. The proposed approach is flexible and general, making it applicable to various objectives and decision-making contexts in the evolving landscape of renewable energy communities. It is applied to two communities with five members, eleven candidate users, multiple preference configurations and a comparison with heuristic metrics from the literature is also addressed. The model also exhibits that equilibrium outcomes and stakeholder behavior are influenced by the order of decisions, their preference criteria, and prospect theory parameters particularly the reference point selection.

%This paper introduces an original approach to an underexplored issue: the integration of a new member into an existing renewable energy community. The problem involves both long-term strategic decisions, such as investment and local pricing, and short-term operational ones, like daily energy and financial flow management. The decision-making process is modeled using extensive-form game theory to account for different time horizons. Specifically, short-term day-ahead scheduling decisions are formulated as a generalized Nash equilibrium problem. Two extensive games are constructed to reflect different sequences of decision-making, enabling analysis of how decision order impacts strategies and subgame perfect equilibria. In the first scenario, an external user is interested in joining a community, representing a situation where the user has limited insight into the community's response. In the second, the community chooses a new member from a set of candidates. These contrasting setups allow exploration of both individual and collective decision-making dynamics. A key strength of the proposed approach lies in its flexibility. Models can be adapted to diverse objectives and can incorporate both perfect and bounded rationality, enabling different settings of prospect theory parameters and reference point selection methods. This adaptability extends models to a wide range of practical and theoretical extension in the evolving landscape of energy communities.
\end{abstract}


\end{frontmatter}

%\newpage

%\tableofcontents

\section{Introduction}

\subsection{Context}
%The growing global demand for energy is placing significant pressure on Europe’s energy systems, which remain heavily dependent on imported fossil fuels. The emergency to advance towards decarbonization has now gained an additional critical dimension: ensuring energy security and independence \lscom{ref?}.
%At the same time, the electricity sector is undergoing a profound transformation, with the rise of distributed energy resources (DERs) and decentralized solutions, such as local solar, small-scale wind generation or individual storage systems. This evolution, driven by both technological progress and increasing environmental and ecological awareness among citizens, has led to the emergence of prosumers—individuals who both produce and consume electricity. %Prosumers have the ability to draw energy from or feed energy into the existing distribution network, actively supporting grid operations while reducing reliance on centralized power plants through self-consumption. 
%The European Union (EU) has recognized the critical role of prosumers in its Clean Energy for all Europeans package \citep{Clean_package19}, which proposes new rules to enhance the flexibility of the electricity system, reduce carbon emissions and empower consumers to play an active and central role in the electricity markets and the decarbonization of the energy system. Among other measures, the package introduces the concept of renewable energy communities (RECs).\par 
%\ppar

Renewable energy communities (\acrshort{REC}s) are new collective actors of the energy system, which gather local consumers and prosumers into organized entities. They may engage in activities such as electricity generation, consumption, supply, storage, aggregation, commercial energy services, sharing and selling energy produced from members' private or community-owned plants \citep{Rossetto23}. They allow their members to gather their energy resources and exchange locally generated renewable energy between participants. In this way, members do not depend solely on traditional wholesale/retail markets structure. Members are free to choose their electricity suppliers for consumption not covered locally and can sell production on conventional markets. \par 
\ppar
Although RECs have attracted growing interest in recent years as an innovative contributor to the energy transition and to a prosumer-centric power system, their implementation has yet to overcome considerable challenges.
This paper explores a crucial aspect that is still not widely studied in the literature: the integration of new members into an existing renewable energy community. The topic is especially relevant to the viability of RECs, which are called upon to expand by integrating members with various characteristics and objectives. European directives \citep{Directive_2018, Directive_2019} state that participation in a REC must be open and voluntary, based on transparent and non-discriminatory criteria. Similarly, any member wishing to leave the community has the right to a fair and non-discriminatory exit procedure. However, there are no further details concerning these procedures. The absence of common standards leaves a number of gray areas, which can give rise to uncertainties and concerns for RECs. More precisely,  the impact of a user's integration or exit on community dynamics is not fully anticipated in the literature. Indeed, entry and exit processes raise specific issues for the stability and efficiency of the REC: the arrival of a new user or the departure of a member can affect energy flows, costs, self-consumption and self-sufficiency rates, etc. In addition, the diversity of user profiles (in terms of consumption, production or flexibility) may require adjustments to strategies and recommendations for energy exchanges and consumption within the REC. To the best of our knowledge, the analysis of the dynamics of community members after a change in their composition (entry or exit), remains relatively unexplored in the literature. This lack of scientific references and regulations is a barrier to the expansion of RECs.

\subsection{Related work}
\citet{MRDP22a} propose heuristic metrics based on community-level needs (generation, consumption, storage) to guide an existing REC to select new members or guide investment decisions. Their approach yields results close to optimization-based methods while offering better explainability and computational efficiency.
\citet{FFP21} develops a community management model with fair revenue-sharing and exit clauses to find the optimal sizing of communities and to enhance cooperation. Users leaving the REC must pay a compensation that decreases over time, to mitigate the impact of their departure. These costs are calculated to ensure that other members do not suffer financial losses, thus guaranteeing economic stability for the REC.
\citet{PA24} consider entry and exit decisions within a bi-level optimization framework: the upper-level determines a new members based on prosumers' preferences for cost and emission savings, while the lower-level maximizes REC's collective welfare, through peer-to-peer trading. These studies highlight the importance of dynamic participation in RECs, but generally rely on heuristic or centralized optimization methods, without explicitly modeling strategic interactions, sequential decisions or behavioral diversity among stakeholders.  
%Once departure and new commitment decisions have been made, the optimal community configuration is recalculated for each period, ensuring that members are well aligned with collective energy and economic objectives.
\par
\ppar 
This paper proposes an original approach to the New Member Integration Problem (NMIP) in an existing REC, where interactions between initial members and a potential newcomer are explicitly modeled. Beyond predefined energy profiles, the candidate may invest in additional generation or energy storage assets, though the focus remains on assessing the overall impact of integration, rather than evaluating specific investment scheme \citep[e.g.,][]{Bauwens19, LO23, SGCGP24}. We model the NMIP over a long-term (\acrshort{LT}) planning horizon while accounting for its repercussions on short-term (\acrshort{ST}) operations. The problem is formulated as a finite extensive-form game, which captures both strategic interactions and the dynamic nature of decision-making. The short-term day-ahead scheduling decisions are modeled as a Generalized Nash Equilibrium Problem (\acrshort{GNEP}), following the framework developed in \citet{SGB25}. While extensive games have been used in power systems, particularly through Stackelberg formulations \citep[e.g.,][]{MZZGB13, LCYZL18, SGVD19, ERESMP19, BPLD20, ABBK21}, their application to the NMIP in RECs with demand-side management (\acrshort{DSM}) schemes \citep{SGB25} remains innovative. Our framework incorporates heterogeneous stakeholder preferences and yields subgames perfect equilibria (\acrshort{SPE}s), providing robust plans of action by eliminating non-credible strategies.\par
\ppar
Standard REC models often assume that prosumers aim to minimize their total costs. However, end-users' growing awareness of current energy and ecological priorities, as well as potential economic, social and environmental benefits that a REC can provide, calls for a broader perspectives. It is increasingly relevant to analyze equilibria when stakeholders show heterogeneous preferences. In this work, we study how the SPEs evolve when the REC and candidates adopt different long-term objectives. For simplicity and readability, our analysis focuses on five criteria, though the framework can accommodate additional objectives, such as those proposed in \citet{MC19, MP19, OBSFDMO22}.
\par
\ppar
Most of the literature assume perfectly rational agents following the \acrfull{EUT} \citep{VNM44}, often with standard risk measures such as expected shortfall or conditional value-at-risk \citep{AHRS18, MPP20, VHK21, LCL21}. However, empirical studies have shown that agent's subjective perception and cognitive biases can significantly impact decision-making under uncertainty and, consequently, final outcomes \citep{SGMP16}. The EUT fails to capture these effects and can not reliably predict actual individual behavior. As an alternative, \citet{KT79} introduced \acrfull{PT} to describe the decision-making of bounded rational individuals under risk. This Nobel-prize-winning theory has already been applied to various problems in power systems. For instance, \citet{WSMP16} and \citet{ESMP18} examine the impact of non-rational behavior on energy management and DSM strategies over multiple time horizons, and \citet{Good19} designs more effective energy policies taking into account cognitive biases and individual preferences. It has also been used in \citet{ERSGMP16, ERESMP19} to show that PT-based preferences influence trading decisions in microgrids, via strategic and Stackelberg games respectively. \citet{DRFAS21} incorporate PT into a REC with a centralized peer-to-peer energy trading, revealing the impact of subjective perceptions on trading decisions and on the overall performance. Recently, \citet{APBP24} propose a local energy market design based on a cooperative game, that combines Shapley-based profit allocation with the representation of non-rational prosumers' behavior. This approach promotes fairness and local self-sufficiency. Although less explored, PT has also been used to analyze investment behavior at household \citep{KD17} and firm level \citep{TMD23}, revealing its relevance in LT decisions. These contributions highlight the importance of modeling individual risk perception, behavioral biases and collective dynamic in energy-related decisions. Our work builds on this literature by integrating PT into a LT REC planning framework, incorporating subjective perceptions in strategic interactions between the REC and candidates.

\par
\ppar
\subsection{Contributions}
This paper addresses the integration of a new member with potential investment into an existing REC via a novel finite extensive-form game framework, grounded in the operational formulation developed in \citet{SGB25}. The main contributions are summarized as follows.
%using the formulation and results previously established in \citet{SGB25}, which provides the theoretical foundation for the operational layer of this study.
\begin{enumerate}[topsep=2pt,itemsep=0pt,parsep=0pt]
   \item We propose an original model combining actions with both long-term consequences (investments, local price adjustments) and short-term operational scheduling ones. The LT decision-making is modeled using finite extensive-form game theory, while ST day-ahead scheduling decisions are formulated as a \acrshort{GNEP}, whose efficient equilibria can be computed through an optimization problem \citep{SGB25}. Our theoretical framework is general and flexible enough to accommodate different decision orders, heterogeneous stakeholder preferences (economic and environmental), and integrates prospect theory to reflect bounded rationality and risk perception of retail import prices in stakeholders' decision-making.
   \item  We apply our \acrfull{NMIP} models to a detailed case study involving two RECs, each composed of 5 members, 11 candidate users and 25 combinations of preference criteria:
   \begin{enumerate}[topsep=2pt,itemsep=0pt,parsep=0pt] 
    \item We compare heuristic methods from \citet{MRDP22a} with subgame perfect equilibria of the game where the REC moves first. We show that heuristic metrics can predict the REC's choice when it adopts financial objectives (e.g., total cost, net present value), except for the price per kWh. As the metrics do not capture environmental aspects, they provide limited insight when the REC prioritizes carbon emissions, underscoring the added value of our framework.
    \item We perform extensive parametric studies and show that equilibrium outcomes and stakeholder behavior are influenced by three key factors: (i) the decision order between the REC and candidates, (ii) preference criteria of both REC and candidates, and (iii) prospect theory parameters, which may induce deviations from the rational benchmark and even counter-intuitive outcomes, especially through the choice of the reference point.

   \end{enumerate}
\end{enumerate}

The remainder of the paper is organized as follows. Section \ref{sec:NMIP} defines the scope and assumptions of the REC and the NMIP where a user is the first mover, along with the time horizon decomposition. The theoretical presentation of extensive-form game formulations, preference criteria and prospect theory are detailed in Section \ref{sec:th-modeling}. Section \ref{sec:NMIP2} details a second NMIP, in which the REC takes the first move. Section \ref{sec:CaseStudy} presents the case-study data. Simulation results for rational and bounded rational stakeholders are discussed in Section \ref{sec:ResRat} and \ref{sec:PTvsEUT}, respectively. The conclusions are reported in Section \ref{sec:Conclusions}.

\section{New member integration problem (NMIP)}\label{sec:NMIP}

%\fbox{Ce serait peut-être plus clair que de ne parler que d'une variante du problème dans les sections 2 et 3, et faire une nouvelle section 4, très brève pour parler de l'autre variante. }

\subsection{Problem scope and hypotheses}\label{subsec:scope}

\begin{figure}[ht!]
    \centering
    \includegraphics[width=0.6\linewidth]{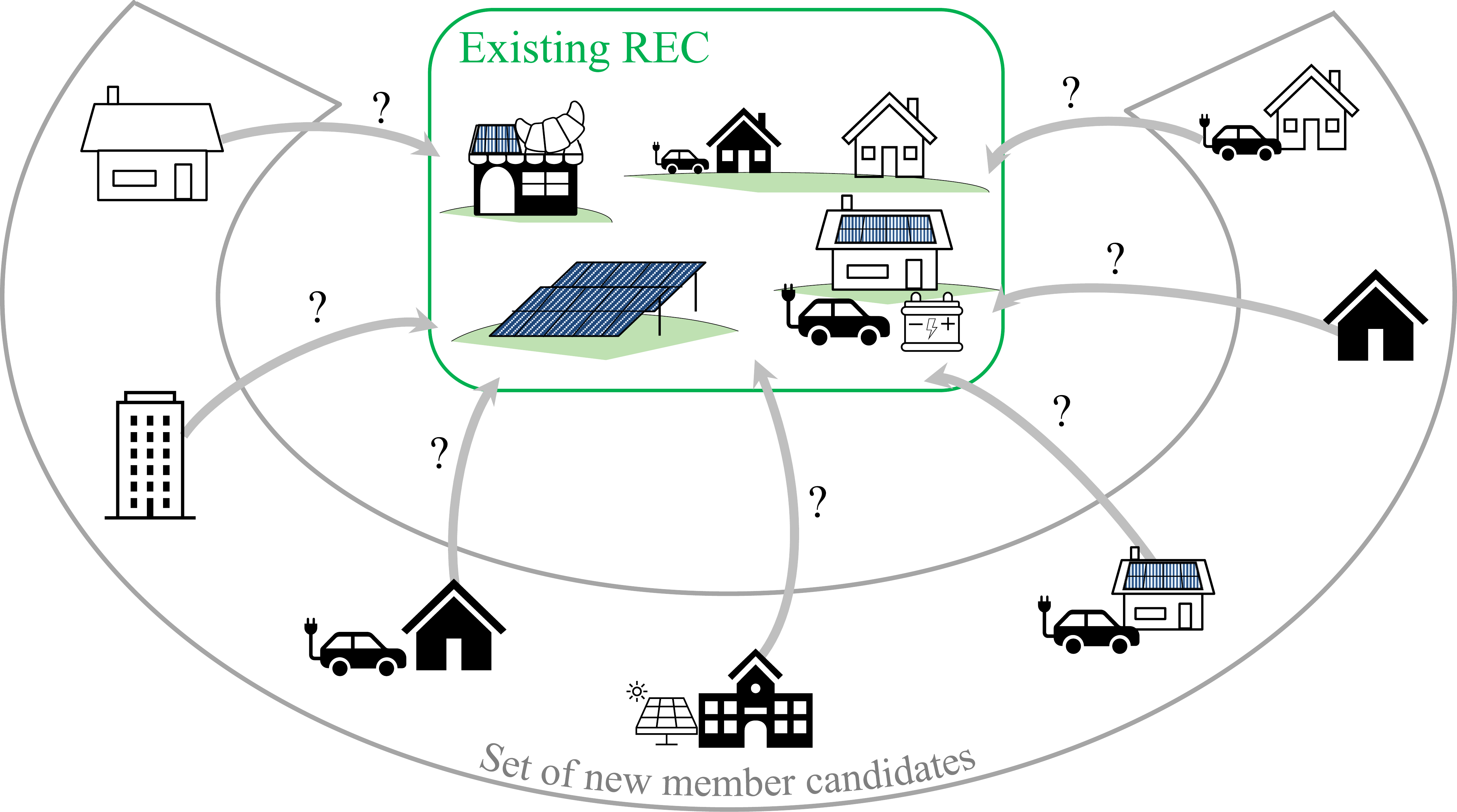}
    \caption{Integration of a new member in an existing REC.}
    \label{fig:NMIP}
    \vspace{-1em}
\end{figure}

The \acrfull{NMIP} is defined for a collaborative community built on a DSM scheme and composed of consumers and prosumers connected to the same \acrshort{LV} distribution feeder (Fig.\ref{fig:NMIP}). The members can purchase their green electricity locally in the REC pool where the excess of local PV productions are mutualized, and from retail markets for gray electricity not produced locally. This corresponds to a REC with design D2 proposed in \citet{SGB25}. Each member can have renewable generation as solar panels or energy storage assets. These devices are fully owned by their users, meaning that the ownership and management of each installation remains at the individual level, and without infrastructure pooling. Only surplus local renewable energy can be made available to other members.

\par
\ppar
We suppose an existing REC, with the set of initial members noted $\users=\lbrace 1, \ldots, N\rbrace$ and the set of new member candidates as $\mathcal{M}=\lbrace 1,\ldots,M\rbrace$. We propose two variants that differ by their decision order: one where the candidate user moves first (joining and investment decisions), and another where the REC leads by selecting a new member among candidates. We focus here on the first structure, while the second is presented in Section \ref{sec:NMIP2}. This analysis thus compares the dynamics of voluntary spontaneous adhesion with those of integration driven by established REC members or the community manager.\par
%We propose two distinct approaches. In the first structure, we model the case of an external user interested in joining the community, with or without investment contribution. The second approach examines the situation where the REC is the instigator of its own expansion
\ppar

%\fbox{TB: il reste une certaines ambiguité dans cette phrase, car on ne sait pas quand sont prises les actions. Mais c'est peut-être trop de donner toutes les infos dans cette première phrase.}

Two time horizons are considered for the new member integration problem, reflecting long-term strategic actions and short-term operational decisions over $Y$ years. \acrlong{LT} decisions correspond to planning actions with LT consequences, i.e., actions taken at time 0 that significantly affect the long run, notably on the daily operational management of energy resources (Section \ref{subsec:LTM}). \acrlong{ST} actions concern day-ahead (\acrshort{DA}) energy resources scheduling, as described in \citet{SGB25}. The NMIP therefore studies the profitability and viability of LT decisions with respect to individual preferences within a REC framework. We initially assume that agents are interested in minimizing their total costs over the $Y$-year period.

%In both approaches, we assume that agents choose actions which will have a significant impact on their daily operational management over the next $Y$ years. \fbox{TB: Remplacer par une phrase plus courte: Two time horizons will be considered.} The new member integration problem is divided into two levels \fbox{TB: Est-ce que level est vraiment un bon mot? On peut juste parler d'actions with long term consequences, et parler de deux types d'actions. Je pense que tu parles de level car tu as déjà le jeu en tête, mais à ce stade, ça ne me semble pas clair.} to account for the different time horizons, up to $Y$ years. The \acrlong{LT} level is associated with planning decisions with LT consequences, i.e., decisions taken at time 0 and which have a significant impact on the long run, notably on the daily operational management of energy resources (Section \ref{subsec:LTM}). The \acrlong{ST} level is dedicated to the day-ahead (\acrshort{DA}) energy resources scheduling decisions as described in \citet{SGB25}. The purpose of the NMIP is, therefore, to study the profitability and viability of LT decisions on individual preferences, within a REC framework. We initially assume that agents are interested in minimizing their total costs over the $Y$-year period.
%To this end, we consider adequate short-term operational management within a REC, considering the short-term decisions of the initial community members and the candidate user.

\subsection{Short-term operational model}\label{subsec:STM}

%\fbox{Je dirai tout de suite que c'est un GNEP, mais que tu vas pouvoir le résoudre comme un problème centralisé grâce à tes résultats précédents.}

\begin{figure}
    \centering
    \includegraphics[width=0.65\linewidth]{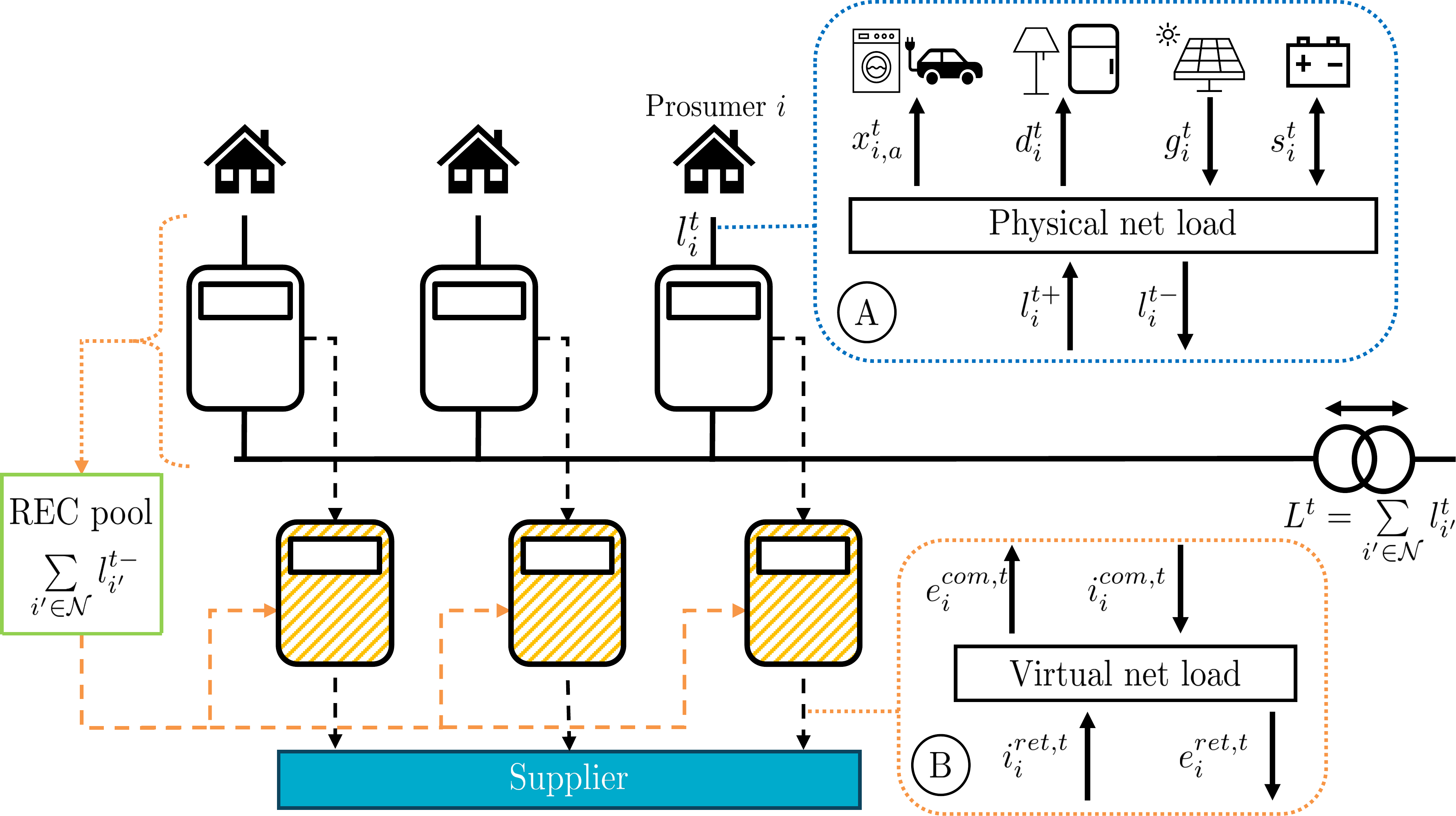}
    \caption{\acrlong{REC} Model \citep{SGB25}}
    \label{fig:REC-ST}
    \vspace{-1em}
\end{figure}

The DSM framework is modeled as a day-ahead energy resources scheduling problem, where the REC's total energy cost is minimized by optimally coordinating the members' resources and exchanges (Fig.\ref{fig:REC-ST}). 

It is formulated as a convex optimization problem:
\vspace{-0.75em}
 \begin{equation}\label{CM}
\underset{\Theta\in \Omega_{\rm{ST}}}{\min}  \ f(\Theta),
\end{equation}
where $\ConstSet_{\rm{ST}}\subseteq\mathbb{R}^n$ is the convex feasible set. Each user $i$ controls a set of decision variables $\Theta_i=\{x_{i,a},s_i,l^+_i, l^-_i, i^{com}_i,e^{com}_i,i^{ret}_i,e^{ret}_i,\overline{p}_i\}$ and we define $\Theta := (\Theta_1,\ldots, \Theta_N)$. The scheduling horizon is discretized into a set of time slots $\mathcal{T}=\lbrace 1,\ldots,T \rbrace$, each of duration $\Delta t$. Each member $i \in \users$ may own flexible devices ($x_{i,a}$), PV panels ($g_i$) and battery storage system ($s_i$). We consider perfect forecasts for both non-flexible demand ($d_i$) and local generation. The physical net load of member $i$ at the time $t$ denoted $l^t_i$, reflects the power exchanged with grid: positive when importing ($l^t_i\geqslant 0 $), and negative when exporting ($l^t_i<0$). We define $l^{t+}_i = \max (0,l^t_i)$ and $l^{t-}_i = \max (0, -l^t_i)$ as the positive and negative net load respectively, such that $l_i^t =l^{t+}_i-l^{t-}_i$. The daily peak power consumption of user $i\in\users$ is denoted by $\overline{p}_i$.\par
\ppar
Virtual flow\footnote{representing commercial, monetary-based flows.} variables are defined for billing purposes. When the net load is positive, energy is imported from the REC pool $i^{com,t}_i$ and/or the supplier $i^{ret,t}_i$. In case of energy surplus (negative net load), it can be sold to other members $e^{com,t}_i$ and/or to the supplier $e^{ret,t}_i$.\par
\ppar

The total cost minimization is subject to constraints reflecting technical, capacity and economic limitations; see \citet{SGB25} for full development details. Among these, global balance constraints ensures that, at each time step, the total energy injected into the REC matches the total energy withdrawn.\par
\ppar

The REC's total costs, for any profiles $\Theta\in\ConstSet_{\rm{ST}}$, are given by: 
%\begin{equation}\label{funct:rec-opcosts}
%\begin{split}
%    f(\Theta)=\sum_{t\in\mathcal{T}}&\Big[\sum_{i\in\users} \lambda^t_{imp}i^{ret,t}_i + \lambda^{t}_{iloc}i^{com,t}_i - \lambda^t_{exp}e^{ret,t}_i - \lambda^t_{eloc}e^{com,t}_i\\
%    &+\alpha (i^{ret,t}_i+\gamma i^{com,t}_i) \Big] + \sum_{i\in\users} \beta \overline{p}_i.
%    \end{split}
%\end{equation}
\begin{equation}\label{funct:rec-opcosts}
    f(\Theta)=\sum_{t\in\mathcal{T}}\Big[\underbracket[0.187ex]{\sum_{i\in\users} \lambda^t_{imp}i^{ret,t}_i + \lambda^{t}_{iloc}i^{com,t}_i - \lambda^t_{exp}e^{ret,t}_i - \lambda^t_{eloc}e^{com,t}_i}_{\text{(C1)}}
    +\underbracket[0.187ex]{\alpha (i^{ret,t}_i+\gamma i^{com,t}_i) \Big] + \sum_{i\in\users} \beta \overline{p}_i}_{\text{(C2)}}.
\end{equation}
The line (C1) represents commodity costs, where the first two terms are the costs of buying energy from the supplier and REC pool. The third and fourth terms show the revenues from selling energy to the supplier and to the REC pool. We assume a single external supplier for all exchanged with the grid. The line (C2) corresponds to grid costs, which are in line with real tariffs applied in Flanders (Belgium, \citet{SGB23}). The first term represents volumetric-based costs with a possible discount $\gamma$ applied for the energy consumed locally, while the second captures capacity-based costs based on the member's peak consumption.
\par
\ppar

The day-ahead resources scheduling problem can also be formulated as a generalized Nash equilibrium problem (GNEP) \citep{FK10, SGB25}. In this way, strategic interactions between members sharing common resources (energy pool and network) can be captured and the privacy-preserving properties of the associated distributed resolution algorithms can be exploited. In this decentralized framework, each member is modeled as a self-interested player who minimizes their individual daily cost function $b_i$ defined in \eqref{funct:CB}, subject to both individual and global (shared) constraints. Therefore, each player's strategy set depends on the strategies of the other members: $\ConstSet_{\rm{ST},i}(\Theta_{-i})$. Given $\Theta_{-i}$ the strategy profile of the rivals, a member $i\in\users$ solves the following problem
\begin{equation}
\mathcal{G}:=
    \begin{cases} \label{GNEP}
         \underset{\Theta_i}{\min} &  b_i(\Theta_i, \Theta_{-i}) \ \ \ \ \ \forall i\in\mathcal{N} \\
         \text{s.t. } & \Theta_i\in \ConstSet_{\rm{ST},i}(\Theta_{-i}).
    \end{cases}
\end{equation}
A strategy profile $\Theta^*$ is a generalized Nash equilibrium (GNE) of the game $\mathcal{G}$, if for all $i\in\users$ and $\Theta_i\in\ConstSet_{\rm{ST},i}(\Theta_{-i})$, we have $b_i(\Theta^*)\leqslant b_i(\Theta_i,\Theta^*_{-i})$.\par
\ppar
We assume that the total cost is distributed among community members continuously at each time step $t\in\mathcal{T}$. The billing of a member $i\in\users$ is defined by 
\begin{equation}\label{funct:CB}
b_i(\Theta) = \sum_{t\in\mathcal{T}}\Big(\lambda^t_{imp}i^{ret,t}_i + \lambda^{t}_{iloc}i^{com,t}_i - \lambda^t_{exp}e^{ret,t}_i - \lambda^t_{eloc}e^{com,t}_i +\alpha (i^{ret,t}_i+\gamma i^{com,t}_i) \Big) + \beta \overline{p}_i
\end{equation}
In the centralized optimization \eqref{CM}, the allocation is computed ex-post, whereas the cost distribution is endogenized in the members' objective functions for the GNEP \eqref{GNEP}. Thanks to the theoretical and empirical results established in \citet{SGB25}, this GNEP can be solved as a centralized optimization problem \eqref{CM}, ensuring the existence of efficients solutions.

\subsection{Candidate-driven NMIP decisions}\label{subsec:LTM}
\begin{figure}[h!]
    \centering
    \includegraphics[width=0.7\linewidth]{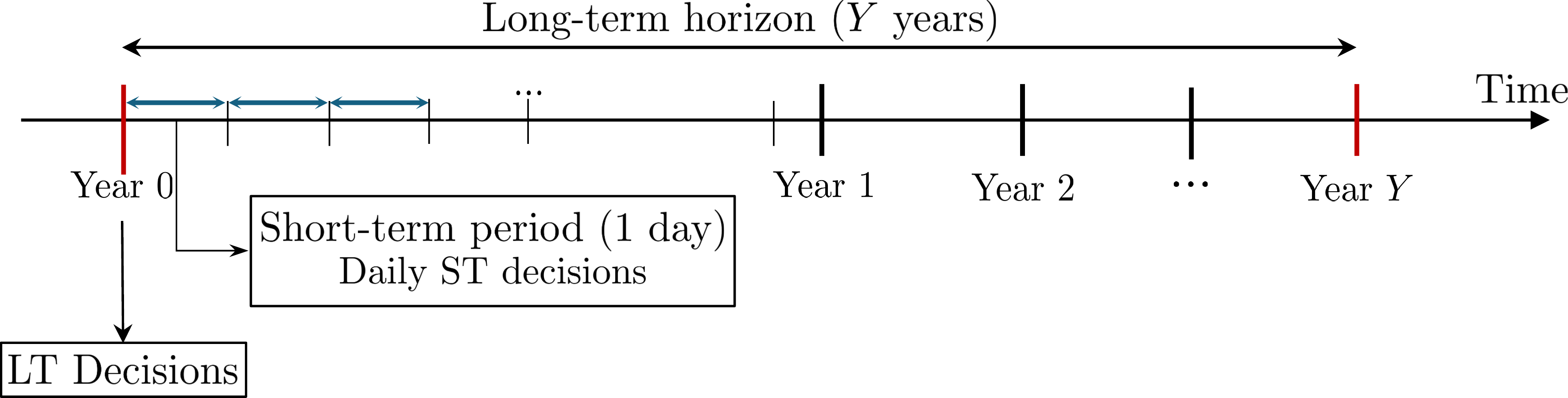}
    \caption{Complete timeline of NMIPs with short-term and long-term periods defined for the various decisions.}
    \label{fig:Timeline}
    \vspace{-1em}
\end{figure}

The new member integration problem considers both LT and ST decisions over a $Y$-year horizon. Long-term actions are taken at time 0, with consequences extending over the entire period. Short-term decisions correspond to daily operational management, as DSM, with a one-day time horizon. Figure \ref{fig:Timeline} shows the complete NMIP timeline and the decision time horizons. In the first instance, we suppose that candidate users and the REC want to minimize their total costs over the $Y$-year period.\par

\ppar

\begin{figure}
\centering
\begin{tikzpicture}[node distance=2cm]
\node (user-1) [Userchoice] {User $j$};
\node (user-2a) [Userchoice, below right of=user-1,xshift=2cm,yshift=2cm] {User $j$};
\node (user-2b) [Userchoice, above right of=user-1,xshift=2cm,yshift=-2cm] {User $j$};
\node (REC-1) [RECchoice, right of=user-2b,xshift=1cm] {REC};
\node (rand1) [Randomechoice, right of=user-2a, xshift=1cm,scale=0.75] {Chance};
\node (rand2) [Randomechoice, right of=REC-1, xshift=1cm,scale=0.75] {Chance};
\node (ST1) [STchoice,right of=rand1,xshift=1cm] {ST scheduling};
\node (ST2) [STchoice,right of=rand2,xshift=1cm] {ST scheduling};
\draw [arrow] (user-1) -- node[anchor=south] {Alone} (user-2a);
\draw [arrow] (user-1) -- node[anchor=north] {Enter} (user-2b);
\draw [arrow] (user-2a) -- node[anchor=south,scale=0.8] {Investment} (rand1);
\draw [arrow] (user-2b) -- node[anchor=north,scale=0.8] {Investment} (REC-1); 
\draw [arrow] (REC-1) -- node[anchor=north,scale=0.8] {Local prices} (rand2);
\draw [arrow] (rand1) -- (ST1);
\draw [arrow] (rand2) -- (ST2);
\end{tikzpicture}
\caption{Schematic decision tree of the candidate-driven NMIP.}\label{fig:Schmtree1}
\vspace{-1em}
\end{figure}
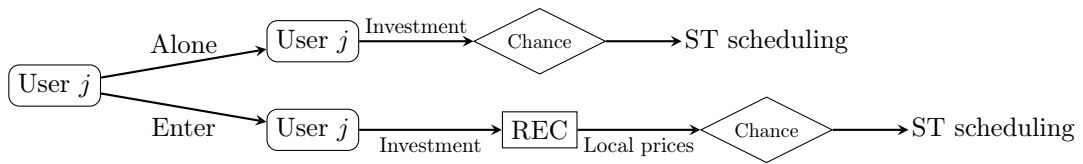

%\fbox{Je pense que ce serait beaucoup plus clair de décrire tout le process de façon informelle avant d'introduire toutes les variables. Tu peux donner l'arbre plus vite si ça aide, même si on ne le formalise que plus tard. La première étape est claire. Pour la deuxième étape, tu peux dire qu'elle choisit un profil d'investissement, sans le décrire dans un premier temps. Pour la troisième étape (communauté), tu dis que la communauté pourra réagit en ajustant ses tarifs, et enfin tu dis qu'on considère des scénarios d'évolution du prix retail. Une fois tout ça expliqué, tu reviens sur la modélisation.}

The Figure \ref{fig:Schmtree1} shows the decision structure of the NMIP where a candidate user moves first. At Year 0, an external user $j\in\mathcal{M}$ considers the value of joining an existing REC. The user $j$ decides whether to remain \textbf{Alone} or to \textbf{Enter} the REC. Then, if joining, the user selects one among several investment profiles, defined by PV and storage options. The REC then has the possibility of adjusting its local import and export tariffs, in order to compensate for this new user integration and her energy profile. Finally, the process incorporates the uncertainty linked to the long-term evolution of retail import prices. These steps represent the set of actions with LT consequences, which are therefore decided sequentially at time 0 in Fig.\ref{fig:Timeline}, before any short-term operation. The final stage, \textit{ST scheduling}, corresponds to the set of daily operational decisions. These ST actions are made simultaneously each day, given the LT configurations previously established. Note that the problem considers the ST decisions of all REC members, while at the LT level, it considers the REC as a single entity making LT integration decisions.\par % In other words, it is an external user $j\in\newusers$ who triggers the integration process, investigating whether joining the REC will benefit her own objectives.
 \ppar

% \tbcom{Je voyais les paragraphe suivants plutôt dans la section suivante. On est déjà en train de modéliser le jeu de façon plus précise. Cela ne me semble pas utile à ce niveau.} \lscom{C'est vrai, mais en même temps dans la même section on formalise le problème court terme (section 2.2). Avis de Zach ?}\\
  Let $j\in\mathcal{M}$ be an external user, potentially owning PV installations of capacity $Q_{pv,j}$ kWp. For the sake of simplicity, and since the model does not attempt to determine the user $j$'s optimal investment, we assume a finite set of investment options, numbered from 1 to $K$, is available and noted $\mathcal{Q}$. For all $k\in\mathbf{\lbrace 1,\ldots,K\rbrace}$, we have $q_k=(q_{pv,k},q_{st,k})\in\mathcal{Q}$ where
 \begin{itemize}[topsep=2pt,itemsep=0pt,parsep=0pt]
     \item $q_{pv,k}$ is a PV capacity in [kWp], such as $q_{pv,k}\in \lbrace0,1,\ldots,q_{pv}^{max}\rbrace$ with $q_{pv}^{max} \in\bb{Z}$.
     \item $q_{st,k}$ is equal to 1 if the profile includes a battery with a capacity of $E^{st}$ in [kWh] and a maximum charging power of $M^{\rm{ch}}$ in [kW], otherwise it is 0.
 \end{itemize}
 We consider the first profile $k=1$ to represent the case where the user does not make any new investments, thus $q_1=(0,0)$. The investment costs (CAPEX) of user $j\in\newusers$ for an investment profile $q_k\in\mathcal{Q}$ is defined as: 
 \begin{equation}\label{funct:Inv-costs}
     C_{inv,j}=\lambda_{pv}q_{pv,k}+ \lambda_{cap}E^{st}+\lambda_{pow}M^{\rm{ch}},
 \end{equation}
 where $\lambda_{pv}$ is the photovoltaic price expressed in [\euro/kWp] and $\lambda_{cap}, \lambda_{pow}$ the price of the storage system linked to the energy [\euro/kWh] and power [\euro/kW] components respectively. Once the user $j\in\newusers$ has decided to join the REC with an investment option, the REC can adjust its local fees ($\lambda_{iloc}^t$, $\lambda_{eloc}^t$) to compensate for the impact of expanding membership. The REC can \textbf{I}ncrease, \textbf{D}ecrease, or leave \textbf{C}onstant its local prices. The user $j$'s LT decision variables set is defined as $\Xi^1_j$, and the one of the REC is denoted by $\Xi^1_{\rm{rec}}$, such as $\Xi^1=(\Xi^1_j, \Xi^1_{\rm{rec}})$. We define the feasible set of LT decisions for this approach by $\ConstSet_{\rm{LT}}^1$.\par
 \ppar

We integrate the uncertainty linked to LT retail import prices $\lambda_{imp}^t$ evolutions over the time horizon, through distinct scenarios with associated probabilities. For the sake of representation, this work assumes an example of a set of three scenarios $\ConstSet_c=\lbrace \Psi_1, \Psi_2, \Psi_3 \rbrace$ defined as follow: 1) the price increases slightly each year; 2) the price increases moderately each year; and 3) a price crisis scenario. Each scenario represents a plausible trajectory for the import price, with an associated probability $(p_1,p_2,p_3)$. Note that the method has no restrictions on the number of possible scenarios, as long as it remains finite. It is possible to extend the problem to infinite cases, but this is not discussed here. For each scenario $\Psi$, we associated a function $\psi$ that models the price $\lambda_{imp}$ over the long-term horizon $\mathcal{Y}=\lbrace 1,\ldots, Y \rbrace$. \par
\ppar

%\fbox{Il y a quelque chose d'étrange par rapport à la structure. Pour moi, ici, on est toujours dans la sous-section "long-term decisions", sous-sous-section Case 1. Je ne m'attends donc pas à trouver de la description short term ici. Revoir la structure et ou la place de certains paragraphes.}
Each ST decision is an action in a ST problem, dependent on production and consumption profiles. This ST problem is solved daily over the entire horizon (see Fig.\ref{fig:Timeline}). According to the actions $\Xi^1_j$ chosen by user $j$, the ST problem gathers one or two day-ahead resources scheduling models. These models are used to provide an estimate of operational costs (OPEX). Given $\Xi^1\in\ConstSet_{\rm{LT}}^1$, the ST problem is defined as follows:
\begin{itemize}[topsep=2pt,itemsep=0pt,parsep=0pt]
    \item If Alone $\in \Xi^1_j$, then daily operations for the user $j$ are modeled by a linear optimization problem, minimizing her own energy bill $f_j$ without being in the community
    \begin{equation}
        f_j(\Theta_j)= \sum_{t\in\mathcal{T}}\Big(\lambda^t_{imp}l^{t+}_j- \lambda^t_{exp}l^{t-}_j + \alpha.l^{t+}_j \Big) + \beta\overline{p}_j.
    \end{equation}
    The REC remains with its initial composition and solves the day-ahead energy resources scheduling model in \eqref{GNEP} at subsection \ref{subsec:STM}.
    \item Otherwise, user $j$ joins the community and can also purchase or sell surplus production from the REC pool. The REC must therefore adjust ST recommendations to reflect its new dynamism. Then, the REC solves the day-ahead energy resources scheduling problem in \eqref{GNEP} for all the $N+1$ members of the set $\users\cup\lbrace \text{user } j\rbrace$, considering the new energy profile of the community.
\end{itemize}
The investment profile chosen by user $j$ also has an impact on the ST problem. The user has a basic energy profile, possibly with an initial PV capacity of $Q_{pv,j}$ kWp. For an investment $q_k\in\mathcal{Q}$, she will possess a total panel capacity of $Q_{pv,j}+q_{pv,k}$ kWp and/or a storage battery. This modifies her energy profile and impacts the daily consumption and production behavior of the user $j$. If this user joins the community, the REC's energy profile must now take into account the new member's profile and its impact on that of the original members $\users$. The REC can adjust local tariffs, which also modifies the basic setting of the ST problem. Finally, market purchase prices $\lambda_{imp}$ for the day are set by the function $\psi$ for the scenario and year in effect. 
\par
\ppar
In conclusion, the ST problem is parametrized by agents' LT decisions and the retail price evolution scenario. Given $\Xi^1\in\ConstSet_{\rm{LT}}^1$, $\Psi \in\ConstSet_c$ and a year $y\in\mathcal{Y}$, the results of the ST problem are $\zeta_j(\Xi^1,\psi(y))$ and $\zeta_{\rm{rec}}(\Xi^1,\psi(y))$, representing respectively the individual daily bill of user $j$ and the one of the initial members of the community defined in \eqref{funct:rec-opcosts} (even if user $j$ has entered).
The total operational cost for the $Y$ years is simply the sum of the daily bills over the entire LT horizon $\mathcal{Y}$:
\begin{align}
    \zeta^{\rm{ST}}_j(\Xi^1,\Psi)&:= \sum_{y=1}^Y\sum_{h=1}^{365} \zeta_j(\Xi^1,\psi(y)), \label{funct:zetaST-j} \\
    \zeta^{\rm{ST}}_{\rm{rec}}(\Xi^1,\Psi)&:= \sum_{y=1}^Y\sum_{h=1}^{365} \zeta_{\rm{rec}}(\Xi^1,\psi(y)). \label{funct:zetaST-rec}
\end{align}
Then, the agents' total costs in the NMIP are defined:
\vspace{-0.5em}
\begin{align}
    &C_{tot,j} =C_{inv,j}+\zeta_{j}^{\rm{ST}} \label{funct:Ctot-j} \\
    &C_{tot,\rm{rec}}=\zeta_{\rm{rec}}^{\rm{ST}}. \label{funct:Ctot-rec}
\end{align}

%\fbox{Je pense que des versions simplifiées des arbres seraient plus parlantes que ce tableau très technique. A ce stade, il est plus important de comprendre la structure des décisions que les différentes variables de décisions.}

\section{Theoretical modeling of the NMIP}\label{sec:th-modeling}

\subsection{Extensive game formulation, analysis and resolution}\label{subsec:GT}
Section \ref{sec:NMIP} identifies the actions available to each candidate user $j\in\newusers$ and to the REC, based on their respective preferences. These decisions have to be taken on different time horizons (Fig.\ref{fig:Timeline}). The integration, investment and price adjustment are made at time 0 with LT effects, while operational choices are made on a daily basis. We adopt noncooperative game theory, which describes and analyzes strategic interactions between different rational selfish agents, who make decisions to optimize their own individual objectives. Since the NMIP involves sequential decision-making, static representations (e.g., normal-form games) are inadequate. To explicitly model timing and decision order, we formulate the NMIP as an extensive-form game \citep{OR94}.

\subsubsection{Extensive games}\label{subsec:Game1}
\begin{figure}[ht!]
    \centering
    %\resizebox{\linewidth}{!}{
    \includegraphics[width=0.75\linewidth]{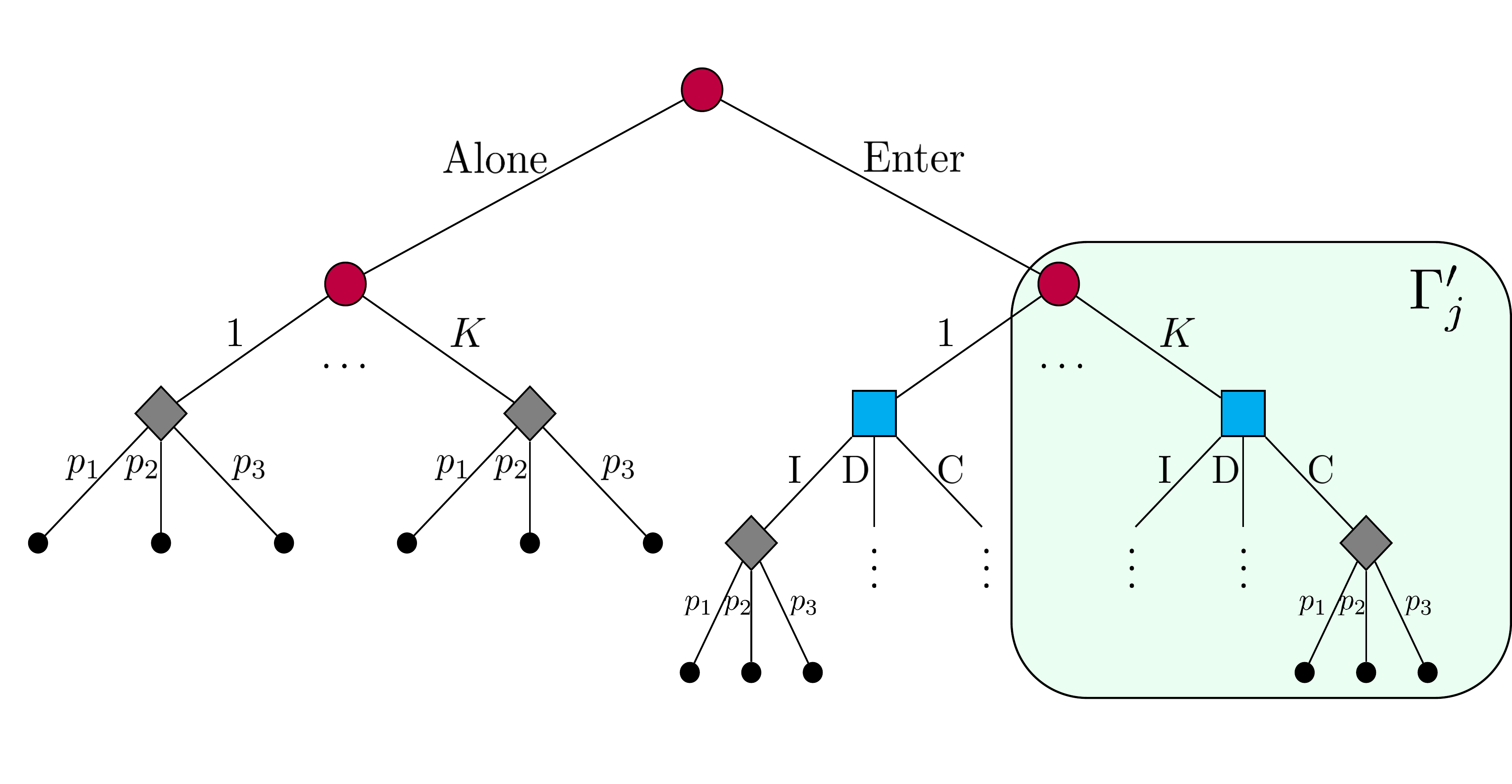}%}
    \par\vspace{0.4cm}
    \includegraphics[width=0.45\linewidth]{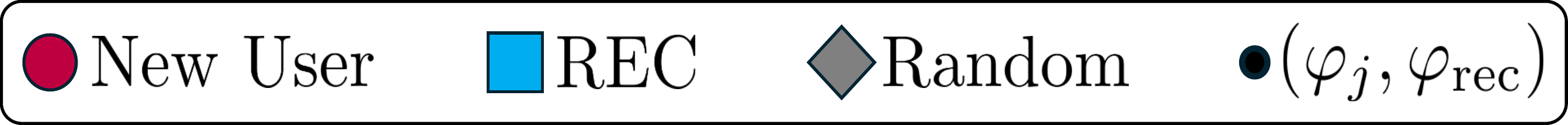}
    \caption{The new member integration game with a new user's point of view.}
    \label{fig:Tree1}
    \vspace{-1em}
\end{figure}

We define the game related to the NMIP when an external user $j\in\newusers$ moves first, described in Subsection \ref{subsec:LTM}. It is a finite extensive-form game with exogenous uncertainty and simultaneous moves in the lower level, structured as a sequential decision tree. For a user $j\in\newusers$, the extensive game $\Gamma_{j}^1$ associated to the NMIP, is represented in Fig. \ref{fig:Tree1} with:
% The first levels of the games can be represented as a tree, which is solved only once
\begin{itemize}[topsep=2pt,itemsep=0pt,parsep=0pt]
    \item The set of players $\mathcal{N}_j^1=\lbrace \text{user }j, \rm{ REC} \rbrace$, and the gray diamond in Fig.\ref{fig:Tree1} corresponds to nodes where \textit{chance} determines the action taken,
    \item The action set $\ConstSet^1_{\rm{LT},c}:=\ConstSet_{\rm{LT},j}^1\times \ConstSet_{\rm{LT},\rm{rec}}^1 \times \ConstSet_c$ such as
    \begin{itemize}[topsep=2pt,itemsep=0pt,parsep=0pt]
        \item The action set of user $j$ is $\ConstSet_{\rm{LT},j}^1:=\lbrace \text{Alone, Enter}\rbrace \cup \mathcal{Q}$,
        \item The REC's actions set is $\ConstSet_{\rm{LT},\rm{rec}}^1:=\lbrace \text{Increase, Decrease, Constant} \rbrace$,
        \item The set of actions available to chance is $\ConstSet_c:=\lbrace \Psi_1, \Psi_2, \Psi_3 \rbrace$.
    \end{itemize}
    \item The set of terminal nodes of the game $\mathcal{Z}^1$,
    \item A set of functions $\varphi^1:\mathcal{Z}^1\to\bb{R}$ assigning total costs to players at each terminal node $z\in\mathcal{Z}^1$.
\end{itemize}
Exogenous uncertainty introduces probabilities associated with the evolution of the market's electricity import price over the $Y$ years. Each branch representing this uncertainty is marked by a probability. For the remainder of this work, we do not explicitly indicate the exponent in the notations for the sake of clarity.

\par
\ppar

The terminal nodes of this tree are associated with a function $\varphi$ that determines the players' total costs (\eqref{funct:Ctot-j}-\eqref{funct:Ctot-rec}) over the $Y$ years. We noted in Section \ref{subsec:LTM} that the total costs include notably operational costs for the $Y$-year period. They are provided by $\zeta^{\rm{ST}}$, which is actually the sum of daily invoices over the entire horizon $\mathcal{Y}$. As a reminder, energy management decisions must be taken on a ST time horizon of one day. For each day of the years studied, daily bills are results of the ST problem defined according to the terminal node $z\in\mathcal{Z}$ and the ongoing year $y\in\mathcal{Y}$. Thus, the user $j$'s individual bill $\zeta_j$ is either the optimal result of \eqref{CM} (if Alone $\in z$), or the outcome valued at a GNE of a GNEP \eqref{GNEP}. Whereas, the total bill of the initial members of the REC $\zeta_{\rm{rec}}$ is the result obtained at a GNE from a GNEP \eqref{GNEP}.
\par
\ppar
In summary, for each terminal node $z\in\mathcal{Z}$, the ST problem %(GNEP and possibly optimization problem) 
is solved daily for each day of the time horizon $Y$, and results are accumulated as each run is completed. In addition to LT decisions, $\zeta^{\rm{ST}}_{j}$ and $\zeta^{\rm{ST}}_{\rm{rec}}$ directly influence the utilities attached to each terminal node $z\in\mathcal{Z}$:
\begin{align}
    &\varphi_j(z)=C_{inv,j}+\zeta_j^{\rm{ST}}(z), \label{funct:userj_Ctot}\\
    &\varphi_{\rm{rec}}(z)= \zeta_{\rm{rec}}^{\rm{ST}}(z),\label{funct:rec_Ctot}
\end{align}
which represents total costs over the entire period.

\par
\ppar
The subgame perfect equilibrium (\acrshort{SPE}) is the relevant solution concept for these games. A strategy profile is a SPE if it constitutes a Nash equilibrium at every decision nodes of the game regardless of prior moves, i.e., each player's strategy must be optimal at every stage \citep{OR94}. In our case, this means that all LT decisions must be optimal given future actions, while satisfying the GNEP equilibria and global minima of linear programs. Due to exogenous uncertainty, each strategy profiles induces a probability distribution over terminal nodes, and players minimize their expected total costs. We denote by $\rm{SPE}(\Gamma)$ the set of SPEs for the game $\Gamma$.

\subsubsection{Equilibrium existence}
We study the SPE existence of a game $\Gamma$. We first ensure that the daily ST problem admits a feasible solution for any LT configuration. The game $\mathcal{G}$ \eqref{GNEP} is modeled as a GNEP under T2 pricing with the continuous allocation method from \citet{SGB25}, which guarantees at least one equilibrium.

\begin{theorem}[\citet{SGB25}]\label{thm:gnep-ex}
Let $\mathcal{G}$ be a GNEP \eqref{GNEP} with the billing function \eqref{funct:CB}. The game has at least one generalized Nash equilibrium.
\end{theorem}
As a direct consequence of Theorem \ref{thm:gnep-ex} applied to the present finite extensive game $\Gamma$, we obtain that, for each terminal node and for every day over the $Y$-year horizon, the short-term problem admits at least one solution. Thus, prior choices and price scenarios only affect the ST models' initial conditions, allowing daily ST problems to be solved independently. The resulting operational costs from successive ST problem solving, are then feed into the cost functions $\varphi$ via $\zeta^{\rm{ST}}$ in \eqref{funct:userj_Ctot} and \eqref{funct:rec_Ctot}. For each user $j\in\newusers$, $\zeta^{\rm{ST}}_j$ in \eqref{funct:zetaST-j}, aggregates individual daily costs, either minimized individually or evaluated at a GNE when joining the REC. Similarly, $\zeta^{\rm{ST}}_{\rm{rec}}$ in \eqref{funct:zetaST-rec} captures the REC's ($\users$) operational costs at a GNE across the horizon. Thus, $\Gamma$ is a finite extensive game with perfect information, conditional on uncertainty. Consequently, we can establish the following result through Kuhn's Theorem \citep{OR94}.

\begin{corollary}\label{cor:exispe}
Given $\Gamma$ the finite extensive-form game with chance and simultaneous moves in Fig.\ref{fig:Tree1}, there always exists a \acrlong{SPE}.
\end{corollary}

\subsubsection{Equilibrium analysis and resolution}
%\fbox{cite tes travaux, sinon c'est trop vague}
Building on theoretical and empirical studies established in \citet{SGB25}, we show that the computation of efficient equilibria in the GNEPs can be reformulated as an optimization problem, which significantly reduces computation time. The GNEPs involved in the ST problems are jointly convex, ensuring the existence of variational equilibria $\rm{VE}(\mathcal{G})$, a subset of GNEs \citep{FK10}. As shown in \citet[Th.5.4]{SGB25}, these \acrshort{VE}s coincide with the social optimum of the centralized problem \eqref{CM}. Moreover, individual bills computed from the centralized solution are nearly identical to those evaluated at the VEs of the GNEP. Hence, it is sufficient to solve the daily linear optimization problem \eqref{CM} to obtained both efficient equilibria and individual costs, so that $\zeta^{\rm{ST}}_{\rm{rec}}$ corresponds to optimal REC costs when no user joins. Each ST problem can be solved independently for each terminal node $z\in\mathcal{Z}$ and scenario $\psi\in\Psi$. Finally, SPEs are determined via backward induction \citep{OR94}, although considering only VE-consistent selections at simultaneous nodes.\par
\ppar

%The following subsections enrich the model's behavioral dimension, while the structural formulation presented above is sufficient to understand the second NMIP in Section \ref{sec:NMIP2} and the subsequent numerical analysis.

\subsection{Modeling heterogeneous preference criteria}\label{subsec:criteria}
Beyond cost minimization, users and RECs may pursue diverse objectives, reflecting heterogeneous preferences (e.g., environmental, social, reliability concerns, etc.).
To better represent this diversity and heterogeneity of motivations, we allow external users and the REC to adopt distinct long-term criteria. We consider the REC as a single entity when making LT decisions, and evaluates each criterion with respect to its original members, even when a new user is integrated.\par
\ppar

Four criteria based on financial indicators as well as one environmental criterion are identified.
\begin{enumerate}%[topsep=2pt,itemsep=0pt,parsep=0pt]
    \item \textbf{Total costs.} The benchmark case involves minimizing the total costs over the time horizon. The aggregate operating costs for the considered period are noted as $C_{op}$. The total costs are
    \begin{equation}\label{funct:Ctot}
        C_{tot}:=C_{inv}+C_{op}.
    \end{equation}
    Note that for the REC, we look at the total operational costs of the REC without the new member to make the LT decisions, then there are no investment costs and $C_{op,\rm{rec}}=\sum_{i\in\users}C_{op,i}$. This is our basic assumption, so in our game $\Gamma^1_j$, functions $\varphi_j$ and $\varphi_{\rm{rec}}$ are defined by \eqref{funct:userj_Ctot} and \eqref{funct:rec_Ctot} respectively.
    \item \textbf{\acrfull{NPV}}. This criterion maximizes the LT profitability of energy investments by integrating the time values of money. The NPV is the discounted sum of all cash-flows associated with an investment project over a period $Y$. For each user $j\in\newusers$: %It is often used to decide whether to accept (NPV $>0$) or reject (NPV$<0$) an investment option.
    \begin{equation}\label{funct:NPV}
    \begin{split}
        \rm{NPV}_j&:= -C_{inv,j} + \sum_{y=1}^{Y} \frac{C_j^{y+}-C_j^{y-}}{(1+\kappa)^y}\\
        &=  -C_{inv,j} - \sum_{y=1}^{Y}\frac{C_{op,j}^y}{(1+\kappa)^y}
    \end{split}
    \end{equation}
    where $C_{inv,j}$ is the investment costs, $C_j^{y+}$ and $C_j^{y-}$ are positive and negative financial flows respectively; and $\kappa$ is the discount rate. In $\Gamma^1_j$, we can consider that each division in the sum of the equation \eqref{funct:NPV} is the result $\zeta_j$ of the ST problem. Thus, $\zeta^{\rm{ST}}_j$ is the sum over the entire LT period $\mathcal{Y}$. Let $z\in\mathcal{Z}$, the payoff function of user $j$ is defined: 
    \begin{equation}
        \varphi_j(z)= -C_{inv,j}- \underbracket[0.187ex]{\sum_{y=1}^{Y}\sum_{h=1}^{365}\underbracket[0.187ex]{\frac{C_{op,j}^{y,h}}{(1+\kappa)^y}}_{\zeta_j(z,y)}}_{\zeta^{\rm{ST}}_j(z)},
    \end{equation}
    with the day $h\in\lbrace1,\ldots,365\rbrace$. Note that the equation is similar for the REC considering only its original members, but with the investment costs equal to 0.
    \item \textbf{\acrfull{ROI}}. It is applied specifically to new users $\newusers$, aiming to maximize the ROI on investments made in the integration process. It is the ratio between net income (over a given period) and investment costs. A high ROI means that investment gains compare favorably with investment costs. It is a way of relating profits to capital invested. For each user $j\in\newusers$ %This is used to evaluate the efficiency of an investments.
    \begin{equation}\label{funct:ROI}
        \rm{ROI}_j=\frac{\text{Net profit of user $j$}}{C_{inv,j}}.
    \end{equation}
    The net profit generated by the investment corresponds to the difference between user $j$'s operational costs in the initial case (i.e., Alone and without investment $q_1$) and those obtained in the case studied $z\in\mathcal{Z}$. Note that we are comparing the elements for the same price scenario realization. Hence, let $\Psi\in\ConstSet_c$, $z_0=$(Alone, $q_1, \Psi$), and $z\in\mathcal{Z}$ such as $\Psi\in z$, the payoff function of user $j$ is:
    \begin{equation}
        \varphi_j(z)=\frac{\zeta^{\rm{ST}}_j}{C_{inv,j}}=\frac{1}{C_{inv,j}}\sum_{y=1}^Y\sum_{h=1}^{365}(\zeta_j(z_0,y)-\zeta_j(z,y)).
    \end{equation}
    \item \textbf{\acrfull{CDE}}. It aims to minimize CDE associated with electrical unit consumption. The daily CDE of the initial members of the REC is defined as follows:%(even if user $j$ decides to join)
    \begin{equation} \label{funct:CDE}
\text{CDE}:=\gamma_{pv}^{CO2}\Big(\sum_{t\in\mathcal{T}}\sum_{i\in\users}i^{com,t}_i\Big) + \gamma_{mix}^{CO2}\Big(\sum_{t\in\mathcal{T}}\sum_{i\in\users}i^{ret,t}_i \Big),
    \end{equation}
    where $\gamma_{pv}^{CO2}$ and $\gamma_{mix}^{CO2}$ measuring greenhouse gas emissions from photovoltaic electricity generation and the national energy mix respectively, in [gCO2eq/kWh]. We then consider that the result $\zeta_{\rm{rec}}$ of a daily ST problem corresponds to \eqref{funct:CDE} evaluated at equilibrium. So, for all $z\in\mathcal{Z}$,
    \begin{equation}
        \varphi_{\text{rec}}(z)=\zeta^{\text{ST}}_{\text{rec}}(z)=\sum_{y=1}^Y\sum_{h=1}^{365} \zeta_{\text{rec}}(z,y).
    \end{equation}
    The reasoning is similar for user $j$. However, if Alone $\in z$, then all the $i^{com}_j$ are equal to zero and $i^{ret}_j$ actually correspond to $l^{+}$ in \eqref{funct:CDE}.
    \item \textbf{Price per kWh [PkWh]}. This criterion aims to reduce the cost of energy per kilowatt-hour to keep energy affordable and accessible to all members
    \begin{equation}\label{funct:PkWh}
        \rm{PkWh}:=\frac{\text{Total operational costs over the entire horizon}}{\text{Positive net charge over the entire horizon}}.
    \end{equation}
    Given the REC, we need to retrieve two daily results in the ST problem. We have $\zeta_{\rm{rec}}=(\zeta_{\rm{rec},1}, \zeta_{\rm{rec},2})$ with $\zeta_{\rm{rec},1}$ the total cost and $\zeta_{\rm{rec},2}$ the aggregated net positive load of the REC members. Then, for $z\in\mathcal{Z}$, the REC's payoff is defined by 
    \begin{equation}
        \varphi_{\rm{rec}}(z)=\frac{\sum_{y=1}^Y\sum_{h=1}^{365}\zeta_{\rm{rec},1}(z,y)}{\sum_{y=1}^Y\sum_{h=1}^{365}\zeta_{\rm{rec},2}(z,y)}.
    \end{equation}
    The case is similar for the user $j$.
\end{enumerate}
 Through the use of these criteria, our method aims to express the economic and environmental motivations of the existing REC, while taking into account the specific preferences of potential newcomers.\par
 \ppar

 Heterogeneity in player preferences only occurs at the global level of NMIPs. For the ST problem, the objective functions used in \eqref{CM} and \eqref{GNEP} remain the same for all players. They consist of minimizing daily total bills. Thus, individual preferences do not modify the dynamics of daily decisions, but they do influence the cumulative results associated with the terminal nodes of the tree, integrating LT results.

\subsection{Modeling the bounded rational behavior using Prospect Theory}\label{subsec:PT}

In this work, the prospect theory is applied exclusively to long-term decision-making under retail import price uncertainty. As a reminder, LT decisions involve evaluating cumulative outcomes at the terminal nodes of the tree, where end-users and RECs face multi-year uncertainty and must consider different alternatives according to their preferences and their perceptions of risks and probabilities. We assume ST decisions are taken rationally, justified by the fact that they may be computed and implemented by a controller without human intervention. Furthermore, since the ST problem scheduling primarily models operational costs across the whole horizon, potential bounded rationality in these decisions is expected to have limited impact on LT outcomes. This justifies the methodological separation from LT and ST analyses.\par
\ppar
%Although we represent the REC as a single entity to simplify long-term modeling, its decisions remain the product of interaction between the members and possibly a community manager, each with their own preferences, objectives and biases. These interactions, combined with often complex collective decision-making procedures (e.g., voting, consensus, delegation) could induce features of bounded rationality on a collective scale. Thus, it would not be astonishing if a REC could adopt behaviors analogous to those observed in individuals. As a result, we consider that the REC can be modeled as a boundedly rational entity and justify the PT framework.\par
%\ppar
Let $\Xi\in\ConstSet_{\rm{LT}}$, we define the set of terminal nodes in the subgame $\Gamma_{|\Xi}$ as
\vspace{-0.5em}
\begin{equation}
    \mathcal{Z}_{|\Xi}:=\set{z\in\mathcal{Z}}{\Xi\sqsubseteq z},
\end{equation}
where $\Xi$ is a prefix of $z$. Prospect theory posits that non-rational individuals do not maximize objective expected utility but instead a subjective global value. The global value of an alternative $z\in\mathcal{Z}_{|\Xi}$ is a combination of two key elements: a subjective value function and a probability weighting function, which together capture behaviors observed in real decision-making. \par
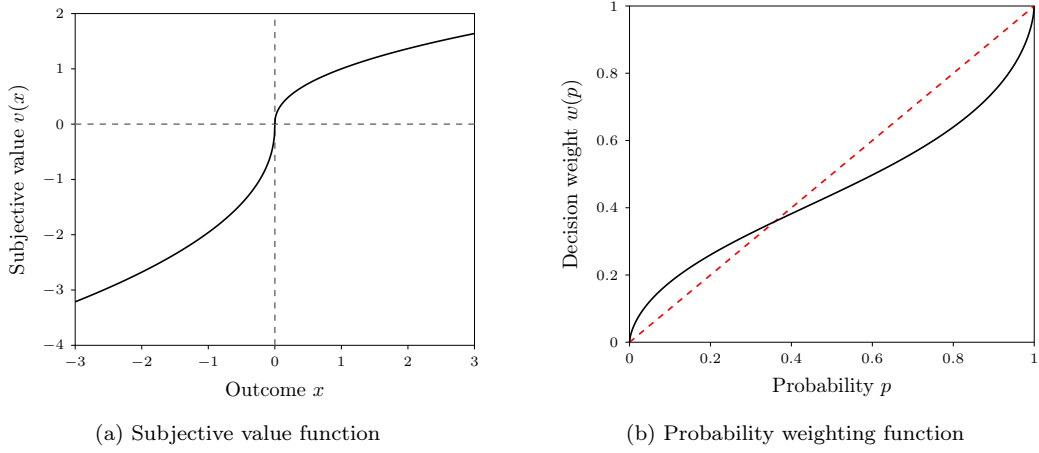
\begin{figure}[ht!]
\centering
\subfloat[Subjective value function]{
\resizebox{0.4\linewidth}{!}{
\begin{tikzpicture}
    \begin{axis}[xlabel={Outcome $x$}, ylabel={Subjective value $v(x)$}, xmin=-3, xmax=3, ymin=-4, ymax=2, xtick={-3,-2,-1,0,1,2,3}, ytick={-4,-3,-2,-1,0,1,2}, tick style={line width=0.5pt,black}, tick label style={font=\scriptsize}, major tick length=2pt, clip=false, samples=500,tick align=outside, ytick pos=left, xtick pos=bottom]
        \def\alpha{0.45} % gain sensibility
        \def\beta{0.45}  % Loss sensibility
        \def\lambda{1.96} % Loss aversion
        \addplot[dashed, gray, thick] coordinates {(0, -4) (0, 2)};
        \addplot[dashed, gray, thick] coordinates {(-3, 0) (3, 0)}; 
        % Function for gains (x >= 0)
        \addplot[domain=0:3,smooth,black,thick]{x^(\alpha)};
         % Function for losses (x < 0)
        \addplot[domain=-3:0,smooth,black,thick]{-\lambda*(-x)^(\beta)};
    \end{axis}
    \end{tikzpicture}
}}
\hspace{0.5cm}
\subfloat[Probability weighting function]{
    \resizebox{0.4\linewidth}{!}{
    \begin{tikzpicture}
    \begin{axis}[xlabel={Probability $p$}, ylabel={Decision weight $w(p)$}, xmin=0, xmax=1, ymin=0, ymax=1, xtick={0,0.2,0.4,0.6,0.8,1}, ytick={0,0.2,0.4,0.6,0.8,1}, tick style={line width=0.5pt,black}, tick label style={font=\scriptsize}, major tick length=2pt, clip=false, samples=500,tick align=outside, ytick pos=left, xtick pos=bottom]
        \def\gamma{0.65} % probability distortion
        \addplot[dashed, red, thick] coordinates {(0,0) (1, 1)}; 
        % weighting function
        \addplot[domain=0:1,smooth,black,thick]{x^(\gamma)/((x^(\gamma)+(1-x)^(\gamma))^(1/\gamma))};
        %intersection: (0.36,0.36)
    \end{axis}
    \end{tikzpicture}}
    }
    \caption{Subjective value function $v$ (left), defined in \eqref{funct:subj-value-funct}, and probability function $w$ (right), defined in \eqref{funct:weighting-funct}. Illustration shown for $r=0$ with parameters $\eta_a=\eta_b=0.45$, $\eta_c=1.96$ and $\eta_d=0.65$.}
    \label{fig:PT-functions}
    \vspace{-1em}
\end{figure}

\ppar
According to PT, outcomes are evaluated relative to a reference point. This is called the framing effect: outcomes above the reference are perceived as gains and those below as losses. Each player has a subjective value function, capturing the subjective perception of outcomes given the reference point. This function should be concave for gains and convex for losses (reflecting diminishing sensitivity) and typically steeper for losses to capture loss aversion. The subjective value function $v:\bb{R}\times\bb{R}\to\bb{R}$ of a player, defined in \citet{KT92} and shown in Figure \ref{fig:PT-functions}.a, is given by:
\begin{equation}\label{funct:subj-value-funct}
    v(-\varphi(z),r):=\begin{cases}
        ((-\varphi(z))-r)^{\eta_a} & \text{if } (-\varphi(z)) \geqslant r\\
        -\eta_c (r-(-\varphi(z)))^{\eta_b} & \text{if } (-\varphi(z))<r,
    \end{cases}
\end{equation}
with $\varphi(z)$ the total cost at the terminal node $z\in \mathcal{Z}$, $r$ is the reference point, and coefficients $\eta_a$, $\eta_b$ and $\eta_c$ control diminishing sensitivity and loss aversion. For preference criteria to be maximized, such as \eqref{funct:NPV}, \eqref{funct:ROI} and \eqref{funct:PkWh}, the subjective value function directly evaluates the outcome $\varphi(z)$ without the minus.\par
\ppar

In this study, the individual reference point represents the initial situation prior to any decision. It reflects the current state in terms of energy asset, associated operational costs, and any other relevant variables. We investigate the impact of the reference point choice on players' strategic behavior in two cases:
\begin{enumerate}[topsep=2pt,itemsep=0pt,parsep=0pt]
    \item \textbf{Fixed reference point} $r^{max}$: defined as the worst-case outcome among the three import retail prices scenarios $\ConstSet_c$ under the initial situation.
    \item \textbf{Stochastic reference point} $r^{stoc}$: scenario-dependent, it captures costs fluctuations in the initial situation due to the import price uncertainty. This approach, introduced in \citet{Sugden03}, has already been applied in energy investment context \citep{TMD23}.
\end{enumerate}
\par
\ppar

The second element is the probability weighting effect. Unlike EUT, where probabilities are treated objectively, empirical studies reveal that individuals tend to overweight low probabilities and underweight high ones. The weighting function reflects these distorted perceptions by assigning subjective decision weights to objective probabilities. It also exhibits diminishing sensitivity: individuals are less sensitive to probability changes near 0 or 1. Based on \citet{KT92} and shown in Fig.\ref{fig:PT-functions}.b, the probability weighting function $w:[0,1]\to[0,1]$ is defined as:
\begin{equation}\label{funct:weighting-funct}
w(p):=\frac{p^{\eta_d}}{(p^{\eta_d}+(1-p)^{\eta_d})^{\frac{1}{\eta_d}}},    
\end{equation}
where $p$ is the objective probability of scenario $\Psi\in\ConstSet_c$ and $\eta_d$ controls the distortion.
%As shown in Figure \ref{fig:PT-functions}.b, the function has an "inverted S" curve with several properties.
\par
\ppar
The global value of an alternative $\mathcal{Z}_{|\Xi}$, given $\Xi\in\ConstSet_{\rm{LT}}$, is computed as:
\begin{equation}\label{funct:PT}
    V(\mathcal{Z}_{|\Xi}):=\sum_{\epsilon=1}^3 w(p_{\epsilon})v(\varphi(z_{\varepsilon}),r).
\end{equation}

In Section \ref{sec:PTvsEUT}, we compare the PT results with the baseline case of EUT, which assume fully rational players maximizing their expected utility:
\begin{equation}\label{funct:EUT}
    U(\mathcal{Z}_{|\Xi}):=\sum_{\epsilon=1}^3 p_{\epsilon}u(\varphi(z_{\varepsilon})),
\end{equation}
where $u:\bb{R}\to\bb{R}$ is the utility function. For simplicity, we assume risk neutrality, i.e., $u(\varphi(z_\epsilon))=\varphi(z_{\epsilon})$.

\section{REC-driven NMIP decisions}\label{sec:NMIP2}
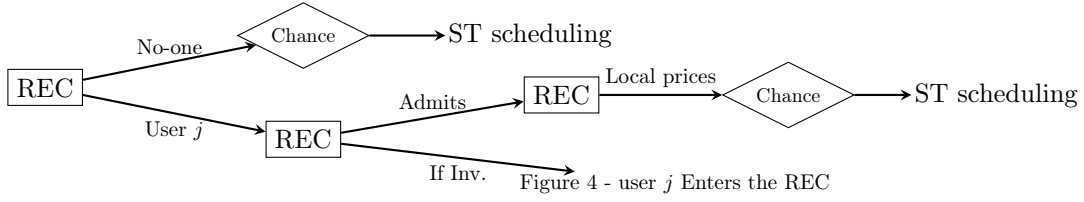
\begin{figure}[ht!]
    \centering
    \begin{tikzpicture}[node distance=2cm]
    \node (REC-1) [RECchoice] {REC};
    \node (rand1) [Randomechoice, below right of=REC-1,xshift=2cm,yshift=2.1cm,scale=0.75] {Chance};
    \node (ST1) [STchoice,right of=rand1,xshift=1cm] {ST scheduling};
    \node (REC-2) [RECchoice,above right of=REC-1, xshift=2cm,yshift=-2.1cm] {REC};
    \node (REC-3) [RECchoice,below right of=REC-2, xshift=2cm,yshift=2cm] {REC};
    \node (rand2) [Randomechoice, right of=REC-3, xshift=1cm,scale=0.75] {Chance};
    \node (ST2) [STchoice,right of=rand2,xshift=0.75cm] {ST scheduling};
    \node (Fig) [STchoice,above right of=REC-2,xshift=3.5cm, yshift=-2cm,scale=0.8] {Figure \ref{fig:Schmtree1} - user $j$ Enters the REC};
    \draw [arrow] (REC-1) -- node[anchor=south,scale=0.8] {No-one} (rand1);
    \draw [arrow] (rand1) -- (ST1);
    \draw [arrow] (REC-1) -- node[anchor=north,scale=0.8] {User $j$} (REC-2);
    \draw [arrow] (REC-2) -- node[anchor=south,scale=0.8] {Admits} (REC-3);
    \draw [arrow] (REC-3) -- node[anchor=south,scale=0.8] {Local prices} (rand2);
    \draw [arrow] (rand2) -- (ST2);
    \draw [arrow] (REC-2) -- node[anchor=north,scale=0.8] {If Inv.} (Fig);
    \end{tikzpicture}
    \caption{Schematic decision tree of the REC-driven NMIP.}\label{fig:Schmtree2}
    \vspace{-1em}
\end{figure}

We now consider the second integration structure, in which the REC takes the lead to its own expansion, as illustrated in Figure \ref{fig:Schmtree2}. The two approaches proposed for integrating a new member into a REC, present significant differences in terms of decision processes and integration perspectives. Then, the set of decisions and the order of decision-making vary according to the approach considered. These differences allow us to analyze how the flexibility or thoroughness of integration processes can influence the REC's dynamics, cost and energy stability, maintaining consistency with its own objectives and end-users' satisfaction.\par
\ppar

The REC has a list of candidates fitting the admission criteria $\newusers$, and evaluates their profiles to identify the one that would best contribute to the stability, energy efficiency and sustainability of the REC in alignment with its objectives\footnote{This paradigm may raise concerns regarding its social implications, as it could limit access for households experiencing energy poverty, potentially reducing inclusiveness within the REC. These aspects are not further discussed here.}. We also consider the case where no candidate on the list is chosen. Hence, the REC selects at most one user among the $M$-candidates. For each candidate $j\in\newusers$, the REC decides whether it \textbf{admits} user $j$ directly, or would agree to accept user $j$ through an \textbf{investment} in new assets (Fig. \ref{fig:Schmtree2}).\par
\ppar

As in Subsection \ref{subsec:LTM}, each candidate can choose from a finite set of investment options $\mathcal{Q}$ and is charged associated costs \eqref{funct:Inv-costs}. Further, except when no one is chosen, the REC can adjust increase, decrease or leave constant local import and export fees ($\lambda_{iloc}^t$, $\lambda_{eloc}^t$). We define the LT decision variables set of a user $j\in\newusers$ and the REC as $\Xi^2_j$ and $\Xi^2_{\rm{rec}}$ respectively, such as $\Xi^2=(\Xi^2_1, \ldots,\Xi^2_M, \Xi^2_{\rm{rec}})$. We denote $\ConstSet_{\rm{LT}}^2$ the feasible set for this approach of the problem. We also take into account the uncertainty of retail import price $\lambda_{imp}$. The rest of the problem description is similar to the first case, so we do not discuss it further.\par
\begin{figure}
    \centering
    \includegraphics[width=0.73\linewidth]{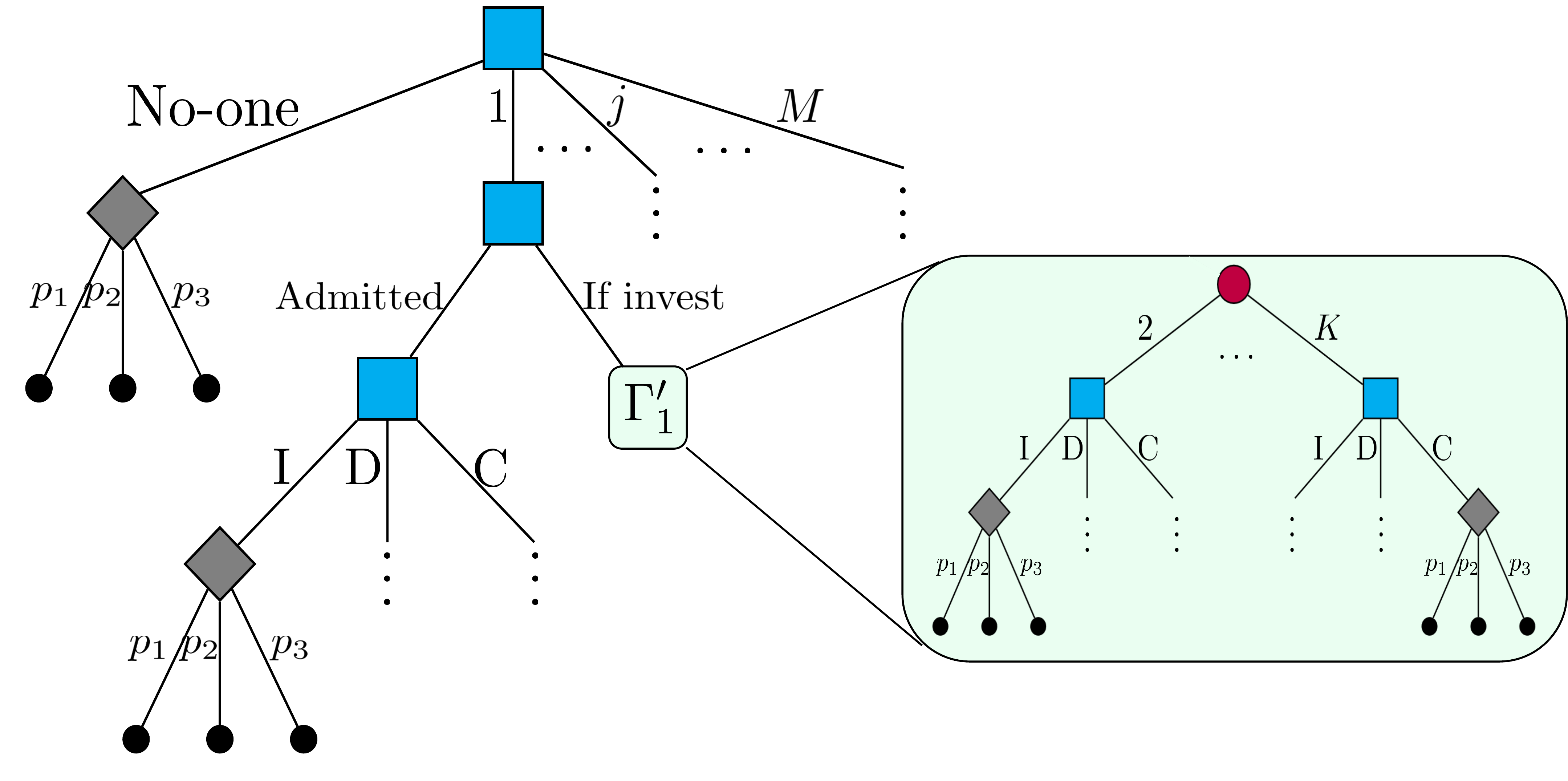}
        \par\vspace{0.4cm}
    \includegraphics[width=0.4\linewidth]{Legend.pdf}
    \caption{The new member integration game when the REC selects a user.}
    \label{fig:Tree2}
    \vspace{-1em}
\end{figure}

\ppar
The extensive game $\Gamma^2$ associated to the NMIP is displayed in Figure \ref{fig:Tree2} with:
\begin{itemize}[topsep=2pt,itemsep=0pt,parsep=0pt]
    \item The set of players $\mathcal{N}^2= \newusers \cup\lbrace \rm{ REC}\rbrace$, again the gray diamonds in Fig.\ref{fig:Tree2} are the nodes where \textit{chance} determines the action taken,
    \item The action set $\ConstSet^2_{\rm{LT},c}:=(\prod_{j\in\newusers} \ConstSet_{\rm{LT},j}^2) \times \ConstSet_{\rm{LT},\rm{rec}}^2 \times \ConstSet_c$ such as 
    \begin{itemize}[topsep=2pt,itemsep=0pt,parsep=0pt]
        \item For all $j\in\newusers$, the action set of user $j$ is $\ConstSet_{\rm{LT},j}^2:=\mathcal{Q} \ (:=(q_k)_{k=1}^K)$,
        \item The REC's set of actions $\ConstSet_{\rm{LT},\rm{rec}}^2:=\newusers\cup \lbrace \text{No-one, Admitted, Invest, I, D, C}\rbrace$,
        \item The set of actions available to chance is $\ConstSet_c:=\lbrace \Psi_1, \Psi_2, \Psi_3 \rbrace$.
    \end{itemize}
    \item The set of terminal nodes of the game $\mathcal{Z}^2$,
    \item A set of functions $\varphi^2:\mathcal{Z}^2\to\bb{R}$ assigning terminal costs to players at each terminal node $z\in\mathcal{Z}^2$.
\end{itemize}
The rest is similar to the first case (Subsection \ref{subsec:Game1}). The theoretical results and methodological elements established in Section \ref{sec:th-modeling} also apply in this second configuration. %(i.e., equilibrium properties, preference criteria and PT)

\section{Case-study}\label{sec:CaseStudy}
We study various NMIP instances over a 20-year horizon ($Y=20$), using a representative year built from 8 typical days (1 weekday and 1 weekend day per season). Two RECs, each composed of 5 members connected behind the same MV-LV feeder, are considered to reflect contrasting electrical profiles (Table \ref{tab:tab-REC-tot}): one with a significant annual energy deficit and one with an annual surplus. 
%\begin{itemize}[topsep=2pt,itemsep=0pt,parsep=0pt]
%    \item A REC with an annual energy deficit, where consumption significantly exceeds local production,
%    \item A REC with an annual energy surplus, where production exceeds consumption.
%\end{itemize}

\renewcommand{\arraystretch}{0.7}
\begin{table}[ht!]
\centering
\scriptsize
%\resizebox{\textwidth}{!}{
\begin{tabular}{lcccccccc}
%>{\columncolor[gray]{0.88}}
\hline
%\rowcolor{lightgray!50}
REC Type & \makecell{PV \\ (kWp)} & ESS & \makecell{Prod. \\ (MWh)} & \makecell{Non-flex. \\ cons. (MWh)} & \makecell{Flex. cons. \\ (MWh)} & \makecell{Total \\ cons. (MWh)} & \makecell{Flex. \\ level (\%)} & \makecell{Cons.-prod. \\ (MWh)}\\ \hline
Deficit & 5 & 2 & 5.851 & 20.351 & 30.384 & 50.735 & 79.75 & 44.884 \\
Surplus & 58 & 3 & 67.873 & 17.811 & 22.824 & 40.635 & 93.38 & -27.238 \\\hline
\end{tabular}%}
\caption{Composition and annual energy profile of 5-members RECs}\label{tab:tab-REC-tot}
\vspace{-1em}
\end{table}

A set of 11 candidates profiles, varying in PV capacity ($Q_{pv,j}$ kWp), energy consumption and flexibility, is considered (Table \ref{tab:tab-newusers}). Two profiles are particularly notable: profile 7 represents a pure supplier with no consumption; and profile 11, specifically designed for the surplus REC (Table \ref{tab:tab-REC-tot}). The latter features non-flexible consumption perfectly aligned with the time steps during which the REC experiences surplus production. Hence, this candidate should theoretically be able to her entire energy needs using the REC's surplus alone. While this makes profile 11 an ideal candidate, it is not a realistic one.\par
\ppar

\renewcommand{\arraystretch}{0.7}
\begin{table}[ht!]
\centering
\scriptsize
%\resizebox{\textwidth}{!}{
\begin{tabular}{lcccccccc}
%>{\columncolor[gray]{0.88}}
\hline
%\rowcolor{lightgray!50}
User & \makecell{PV \\ (kWp)} & ESS & \makecell{Prod. \\ (MWh)} & \makecell{Non-flex. \\ cons. (MWh)} & \makecell{Flex. cons. \\ (MWh)} & \makecell{Total \\ cons. (MWh)} & \makecell{Flex. \\ level (\%)} & \makecell{Cons.-prod. \\ (MWh)}\\ \hline
1 & 0 & 0 & 0 & 25.98  & 0 & 25.98 & 0 & 25.98   \\ 
2 & 10 & 0 & 11.702 & 1.084 & 11.88 & 12.964 & 91.64 & 1.262 \\ 
3 & 0 & 0 & 0 & 1.084 & 0 & 1.084 & 0 & 1.084   \\ 
4 & 10 & 0 & 11.702 & 25.98 & 9 & 34.98 & 25.73 & 23.278 \\ 
5 & 0 & 0 & 0 & 5.473 & 6.552 & 12.025 & 54.5 & 12.025 \\ 
6 & 1 & 0 & 1.17 & 9.933 & 8.352 & 18.285 & 45.68 & 17.115 \\
7 & 20 & 0 & 23.405 & 0 & 0 & 0 & 0 & -23.405 \\ 
8 & 7 & 0 & 8.192 & 2.89 & 5.112 & 8.002 & 63.88 & -0.19 \\ 
9 & 4 & 0 & 4.681 & 2.907 & 0.720 & 3.627 & 19.85 & -1.054 \\
10 & 5 & 0 & 5.851 & 2.383 & 10.44 & 12.823 & 81.41 & 6.972 \\ 
11 & 0 & 0 & 0 & 46.928 & 0 & 46.928 & 0 & 46.928  \\\hline
\end{tabular}%}
\caption{Annual profiles of new user candidates.}\label{tab:tab-newusers}
\vspace{-1em}
\end{table}

Each candidate $j\in\newusers$ faces $K=22$ LT investment options. These include PV installation from 0 to 10 kWp (1 kWp steps) and/or a domestic battery with a capacity of 14 kWh and 5 kW power. Users may also not invest ($q_1$). The photovoltaic price is fixed at $\lambda_{pv}=1500$\euro/kWp, while the tariffs for the ESS are $\lambda_{cap}=300$\euro/kWh and $\lambda_{pow}=300$\euro/kW. As part of LT decisions, the REC may adjust local tariffs $\lambda_{iloc}$ and $\lambda_{eloc}$, by $\pm 0.01$\euro/kWh or leave them unchanged. Three scenarios are considered for the evolution of retail import prices over the next 20 years: a slight increase of 0.005 \euro/kWh per year ($\mathbf{\Psi_1}$), a moderate increase of 0.01 \euro/kWh per year ($\mathbf{\Psi_2}$) and a dynamic scenario following Belgian market trends during the energy crisis\footnote{\samepage based on formulas used to calculate Engie's import tariffs \citep{Engiea}, which vary month by month with the EPEX DAM index \citep{Engieb}.} ($\mathbf{\Psi_3}$). Their respective probabilities are $p_1=1/6$, $p_2=1/3$ and $p_3=1/2$ (see \citet{Zenodo26} for details). The trees of $\Gamma^1_j$ (Fig.\ref{fig:Tree1}) and $\Gamma^2$ (Fig.\ref{fig:Tree2}) contain 264 and 2181 terminal nodes, respectively.\par
\ppar

%We assume three scenarios for the evolution of retail import prices over the next 20 years:
%\begin{itemize}[topsep=2pt,itemsep=0pt,parsep=0pt]
%    \item $\mathbf{\Psi_1}$: The price increases by 0.005 \euro/kWh each year, 
%    \item $\mathbf{\Psi_2}$: The price increases by 0.01 \euro/kWh each year,
%    \item $\mathbf{\Psi_3}$: The price follows the electricity price trends on the Belgian market during the energy crisis\footnote{based on formulas used to calculate Engie's import tariffs \citep{Engiea}, which vary month by month with the EPEX DAM index \citep{Engieb}.}
%\end{itemize}
%with their associated probability $p_1=1/6$, $p_2=1/3$ and $p_3=1/2$ (see \lscom{Zenodo} for details). 

%\begin{figure}
%    \centering
%    \includegraphics[width=0.58\linewidth]{Scenar}
%    \caption{Import retail price evolution.}
%    \label{fig:scenar}
%\end{figure}

Non-flexible consumption profiles are extracted from the \citet{Pecan} dataset, with $T=24$ hourly time steps and assumed unchanged over the simulation horizon. Community members and candidate users may own different flexible devices: dishwashers, washing machines, clothes dryers, electrical vehicles, and heat pumps with seasonal heating needs. The initial and final battery state-of-charge are fixed at 50\%. Bi-hourly tariffs are applied for imports, exports, and internal exchanges, varying between night (21:00–04:00) and daytime. Full profiles and tariff parameters are detailed in \citet{Zenodo26}.\par
\ppar

Building the extensive games $\Gamma^1_j$ and $\Gamma^2$ involve solving 42,720 and 354,240 optimization problems, respectively. Short-term problems are implemented in Julia 1.8 using the JuMP package, and solved using Gurobi. Extensive games are modeled in Python 3.11.6 using the nutree library. Terminal nodes store the players' gains or costs, computed from the ST simulations based on the chosen preference criterion. Subgame perfect equilibria are determined using backward induction, starting from terminal nodes up to the root. Computations were run on an Intel(R) Core(TM) i7-1260P 2.10 GHz with 325 Go of RAM. The whole range of ST simulations for a REC and a candidate user from $\newusers$, were solved at most 147.20s.
\par
\ppar
We tested several combinations of preference criteria from Section \ref{subsec:criteria}: total costs \eqref{funct:Ctot}, NPV \eqref{funct:NPV}, CO2 emissions \eqref{funct:CDE}, ROI \eqref{funct:ROI} and price per kWh \eqref{funct:PkWh}. For the NPV, we consider two discount rates based on \citet{Statbel}: $\kappa_1$=4.14\% (NPV1, July 2023) and $\kappa_2$=0.36\% (NPV2, October 2023). We assume that candidate users adhere to the same preference; heterogeneity arises between $\newusers$ and the REC. In total, 25 game-user combinations were simulated $j\in\newusers$. For conciseness, we focus on the most relevant results. %The initial values of the various criteria are shown in Table \ref{tab:RECs_init-values}.
\begin{comment}
\renewcommand{\arraystretch}{0.7}
\begin{table}[ht!]
    \centering
    \scriptsize
    \
    \begin{tabular}{lcc}
    \hline
         & Deficit & Surplus  \\
         \hline
         NPV1 & -144 253.34\euro & -43 524.35\euro \\
         NPV2 &  -212 943.88\euro & -65 448.51\euro \\
         $C_{tot}$ &  221 640.703\euro & 68 231.05\euro \\
         CDE & 137.564 tCO2eq & 69.016 tCO2eq \\
         PkWh & 0.239\euro/kWh & 0.147\euro/kWh \\
         \hline
    \end{tabular}
    \caption{Expected utilities $U_{\rm{rec}}$ of existing RECs without new members for each criterion.}
    \label{tab:RECs_init-values}
    \vspace{-1.5em}
\end{table}
\end{comment}

\section{Results for perfectly rational agents}\label{sec:ResRat}
This section analyzes results obtained from simulations under the assumption that all agents behave rationally, in line with the EUT \citep{VNM44}, serving as a baseline for evaluating decisions made in our models. Subsection \ref{subsec:heu-vs-spe} observes whether the heuristic methods proposed by \citet{MRDP22a} can predict the candidate selected by the REC in the NMIP modeled as extensive game $\Gamma^2$ (Fig.\ref{fig:Tree2}). Subsection \ref{subsec:ResSPE-order-Rat} investigates how the order of decision-making affects SPE outcomes and stakeholder behavior. Finally, we compared extensive games $\Gamma^2$ (Fig.\ref{fig:Tree2}) and $\Gamma^1$ (Fig.\ref{fig:Tree1}) from the candidates' perspective.

\subsection{Heuristic vs equilibrium}\label{subsec:heu-vs-spe}
The results of the heuristic methods proposed by \citet{MRDP22a} are shown and compared to the SPE outcomes of the extensive game $\Gamma^2$ (Fig.\ref{fig:Tree2}) in which the REC selects one candidate to join. The heuristic approach relies on two metrics: the matching score (\acrshort{MS}), which quantifies how well a candidate’s net load profile compensates for the REC’s energy surplus or deficit; and the collective self-consumption (\acrshort{CSC}), which measures the improvement in local energy use after integrating the candidate. Both metrics are computed from LT energy profiles and use to discriminate users candidates to join an existing REC. For a consistent comparison, we normalize each metric by dividing a user's contribution by the highest observed value.

\subsubsection{REC in deficit}
\begin{figure}[ht!]
    \centering
    \includegraphics[width=0.55\linewidth]{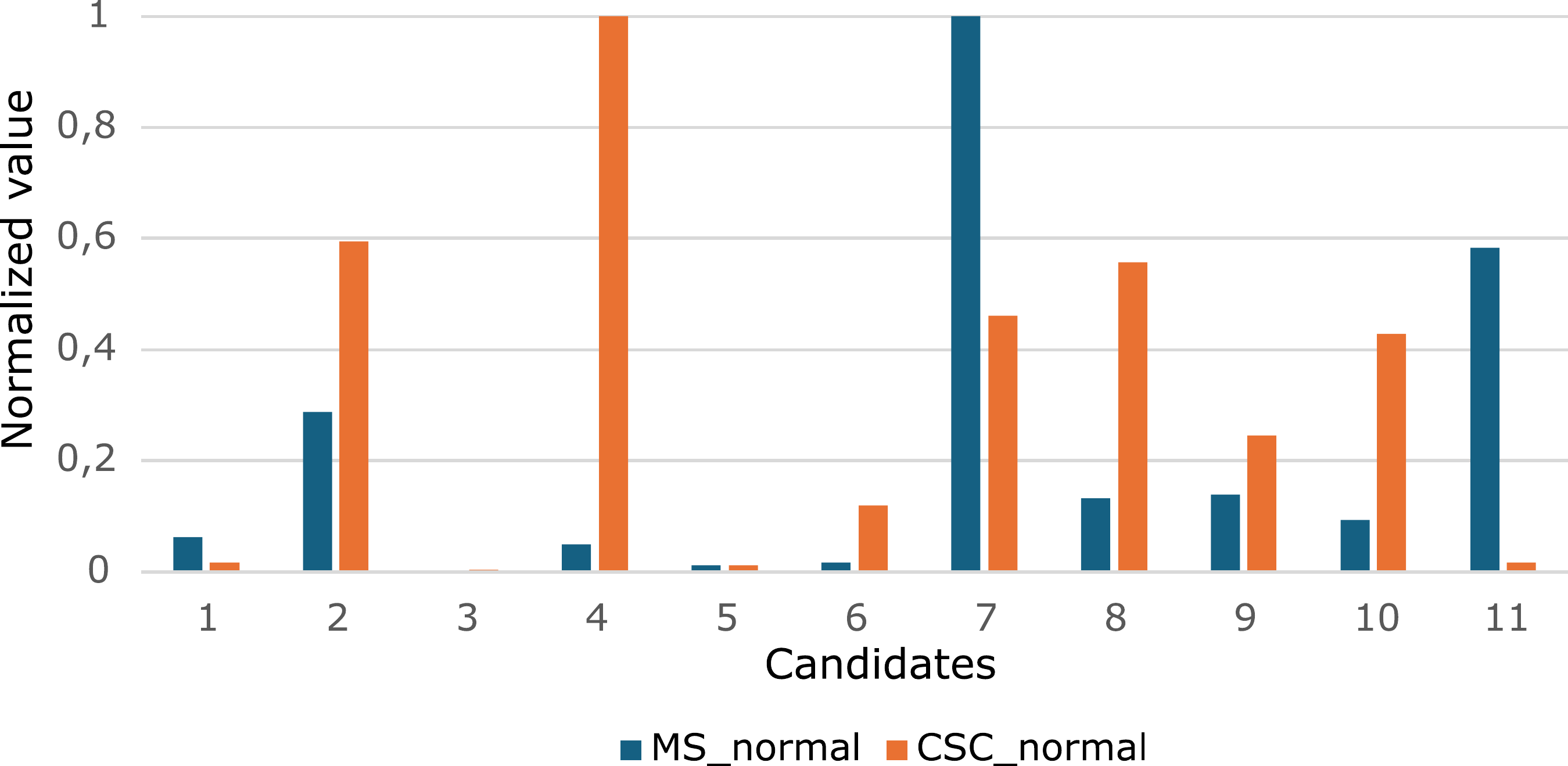}
    \caption{Normalized values for both metrics for each user in $\newusers$ and the energy-deficit REC.}
    \label{fig:Metric_Def}
    \vspace{-1em}
\end{figure}

We study the REC operating under an annual energy deficit (Tab.\ref{tab:tab-REC-tot}), where the reference CSC with five members is 5.677 MWh. The normalized contribution of each candidate $j\in\newusers$ according to the heuristic metrics MS and CSC, are shown in Figure \ref{fig:Metric_Def}. 
\begin{comment}
The descending rankings differ depending on the metric, as detailed below: 
\begin{align}
&\rm{MS}_{\rm{def}}:\lbrace 7, 11, 2, 9, 8, 10, 1, 4, 6, 5, 3\rbrace, \label{metric:MS-def} \\
&\rm{CSC}_{\rm{def}}:\lbrace 4, 2, 8, 7, 10, 9, 6, 1, 11, 5,3 \rbrace.\label{metric:CSC-def}
\end{align} 
\end{comment}
Profiles 1, 3, 5 and 11 are pure consumers, and their contributions are higher under the MS than with CSC in this case. In contrast, profiles 2, 4, 8 and 10 exhibit high individual self-consumption, which is valued in the CSC metric, but not in the MS. User 4 remains a heavy consumer overall (Tab. \ref{tab:tab-newusers}). According to both metrics, users 5 and 3 appear well-suited for this REC. Profile 7, being a large producer, naturally achieves the highest MS in the energy-deficit context, as he can make the production available to other members. However, part of this production could be exported to retail markers. \par
\ppar
We examine the outcomes of the extensive game $\Gamma^2$ in Fig.\ref{fig:Tree2}. Across all tested preference configurations, the REC consistently selects the pure producer candidate 7, regardless of other actions. This predictable result seems to confirm the model's ability to identify the most logical and optimal choice. Given this dominant profile, we exclude user 7 from the candidate list in the remainder: $\newusers'=\newusers\backslash \lbrace 7 \rbrace$. Given the uncertainty in retail import prices, Table \ref{tab:RatSPE-def-tree2} shows the expected utilities $U$ \eqref{funct:EUT} of rational agents across varied preference combination. Despite the diversity of actions, candidate 2 is consistently preferred due to its profile (Tab.\ref{tab:tab-newusers}).
\par
\ppar

\renewcommand{\arraystretch}{0.7}
\begin{table}[ht!]
    \centering
    %\resizebox{\textwidth}{!}{
    \scriptsize
    \begin{tabular}{l|ccccccccc}
    \hline
         &\multicolumn{2}{c}{Criteria}& Num. & \multirow{2}{*}{User} & PV & \multirow{2}{*}{Stor.} & Local & Exp. utility & Exp. utility \\
         &Users $\newusers$ & REC & SPE & & (kWp) & & prices & user & REC \\
         \hline
         1 & NPV1 & NPV1 & 1 & 2 & +0 & 0 & D & -21 071.6\euro & -134 319\euro \\
         2 & NPV2 & NPV2 & 1 & 2 & +0 & 0 & D & -31 328.9\euro & -197 895.7\euro\\
         3 & $C_ {tot}$ & $C_{tot}$ & 1 & 2 & +0 & 0 & D & 32 628.88\euro & 205 942.1 \euro\\
         4 & CDE & CDE & 1 & 2 & +10 & +1 & D & 18.556 tCO2eq & 109.65 tCO2eq  \\
         5 & ROI & (financial) & 1 & 2 & +0 & 0 & D & 0\% & \\
         6 & ROI & CDE & 1 & 2 & +2 & 0 & I & 0.764\% & 119.91 tCO2eq \\
         7 & NPV1 & PkWh* & 243 & 2 & +1 & 0 & I & -20 007.2\euro & 0.216\euro/kWh \\
         8 & NPV2 & PkWh* & 243 & 2 & +0 & 0 & I & -28 982.94\euro & 0.219\euro/kWh \\
         9 & $C_{tot}$ & PkWh* & 243 & 2 & +0 & 0 & I & 30 191,1\euro & 0.219\euro/kWh \\
         10& NPV1 & CDE & 1 & 2 & +2 & 0 & I & -20 565.1\euro & 119.91 tCO2eq \\
         11& NPV2 & CDE & 1 & 2 & +2 & 0 & I & -29 240.79\euro & 119.91 tCO2eq \\
         12& $C_{tot}$ & CDE & 1 & 2 & +0 & 0 & C & 31 410.6\euro & 123.56 tCO2eq\\
         13& CDE & (financial) & 1 & 2 & +10 & +1 & D & 18.556 tCO2eq &  \\
         14& PkWh* & (financial) & 3 & 2 & +10 & +1 & D & 0.12 \euro/kWh & \\
         15& PkWh* & CDE & 6 & 2 & +9 & 0 & I & 0.115\euro/kWh & 110.073 tCO2eq\\
         \hline
    \end{tabular}%}
    \caption{Expected utilities at SPEs of the NMIP $\Gamma^2$ for the REC in deficit with the candidates set $\newusers'$ and the rationality of agents is assumed. The (financial) notation indicates that the results are valid if one of the 3 criteria: NPV1, NPV2 and $C_{tot}$ is used. We have used the lexicographical order in the case of PkWh.}
    \label{tab:RatSPE-def-tree2}
    \vspace{-1em}
\end{table}

The REC accepts user 2 without additional investment and lower local prices under homogeneous financial scenarios 1-3 \& 5. In case 4, with increased production and storage, user 2 can increase self-consumption or inject surplus into the REC pool, thus reducing CO2 from retail imports. It is therefore justifying the REC's interest to demand more investment. To manage complexity, a lexicographic order \citep{OR94} was applied for agents with the PkWh criterion \eqref{funct:PkWh}. Despite this, 243 SPEs arise when users have a financial criterion (NPV1, NPV2 and $C_{tot}$) and the REC minimizes the price per kWh (cases 7-9), though SPE outcomes and actions are identical. Expected values of users 2 are better than in the homogeneous cases 1-3.
Difference between NPV1 and NPV2 highlight the discount rate's impact on decisions. When the REC minimizes the CDE (cases 10-12), user 2's expected utilities surpass homogeneous cases 1-3, but are lower than in the case where the REC follows PkWh criterion. For scenario 12, price remain constant, explaining distinction from case 3. Given lexicographical order for users, we have 3 SPEs with identical outcomes and actions, where user 2 invests in 10 kWp and storage in case 14; similarly in case 15, 6 SPEs occur with a 9 kWp investment yielding better outcomes.\par
\ppar
According to both metrics, user 2 is the preferred candidate after user 7. Then, the selection of user 2 is in line with the expectations given the characteristics of the profile and aligns with the underlying principles of heuristic methods. However, we observe an unusual case where one candidate proves suitable across all preference combinations.

\subsubsection{REC in surplus}
\begin{figure}[ht!]
    \centering
    \includegraphics[width=0.48\linewidth]{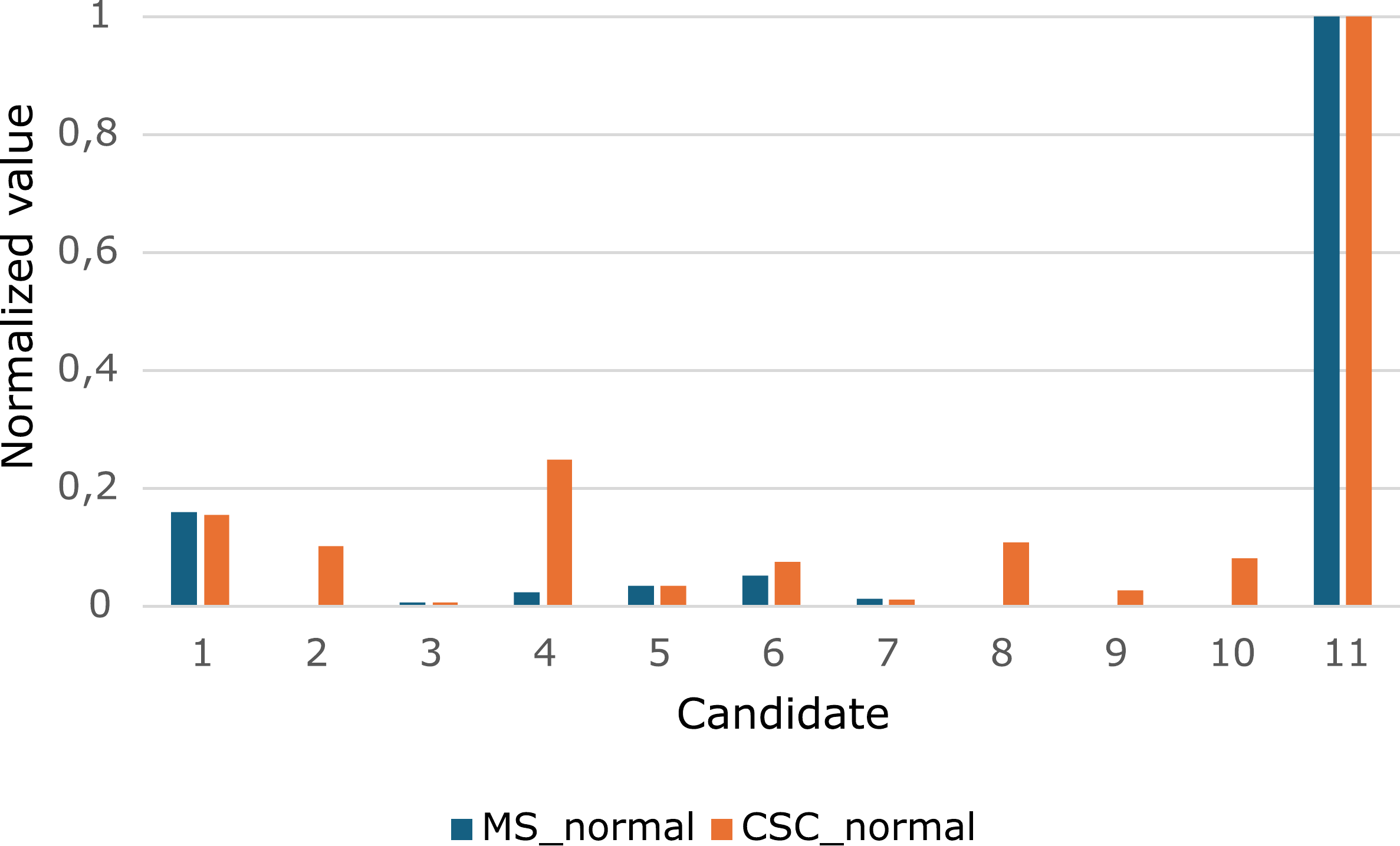}
    \caption{Normalized values for both metrics for each user in $\newusers$ and the energy-surplus REC.}
    \label{fig:Metric_Sur}
    \vspace{-1em}
\end{figure}

We now analyze the REC in a surplus situation (Tab.\ref{tab:tab-REC-tot}), with an annual CSC of 13.663 MWh. The results of both metrics, normalized in respect to candidate 11 who achieves the highest values, are shown in Figure \ref{fig:Metric_Sur}.

\begin{comment}
The rankings in descending order of MS and CSC, are provided by
\begin{align}
&\rm{MS}_{\rm{sur}}: \lbrace 11, 1, 6, 5, 4, 7, 3, 10, 9, 2, 8\rbrace, \label{metric:MS-sur} \\
&\rm{CSC}_{\rm{sur}}: \lbrace 11, 4, 1, 8, 2, 10, 6, 5, 9, 7, 3 \rbrace. \label{metric:CSC-sur}
\end{align}
\end{comment}
User 11 stands out as the optimal candidate for the energy-surplus REC. This profile was specifically designed: the non-flexible consumption aligns precisely with excess production periods, allowing full use of the REC's surplus. Although profile 1 is a large and inflexible consumer, part of her demand is aligned with surplus hours, resulting in a high MS. Excluding candidate 11, user 1 would likely become the preferred option.Once again, profile 3 ranks low due to her limited consumption, which lowers both MS and CSC scores. Adding storage could be a possible solution to improve the performance. Rather than comparing all users uniformly, it may be more insightful to consider candidate groupings based on shared characteristics. As in the deficit case, profiles 2, 4, 8 \& 10 exhibit high individual self-consumption, which is reflected in the CSC, but not in the MS. However, their contribution to the REC's collective use of surplus remains limited. As such their elevated CSC may overstate their actual value to the REC, slightly biasing the CSC ranking.\par
\ppar

We examine the outcomes of the extensive game $\Gamma^2$. The REC systematically selects producer 7 to minimize its CDE, and user 11, tailor-made to meet this REC's specific needs. These results confirm that our model yields consistent and expected actions in terms of user selection. Given their specific particular characteristics, we exclude them from the candidate list for the remainder of the analysis: $\overline{\newusers}=\newusers\backslash \lbrace 7,11 \rbrace$. Table \ref{tab:RatSPE-sur-tree2} reports the actions and expected utilities $U$ \eqref{funct:EUT} associated with the SPEs. To avoid combinatorial explosion of backward induction, we use the lexicographical order for the individual with the PkWh* \eqref{funct:PkWh}.
\par
\ppar

\renewcommand{\arraystretch}{0.7}
\begin{table}[ht!]
    \centering
    \scriptsize
    %\resizebox{\textwidth}{!}{
    \begin{tabular}{l|ccccccccc}
    \hline
         &\multicolumn{2}{c}{Criteria}& Num. & \multirow{2}{*}{User} & PV & \multirow{2}{*}{Stor.} & Local & Exp. utility & Exp. utility \\
         &Users $\newusers$ & REC & SPE & & (kWp) & & prices & user & REC \\
         \hline
         1 & NPV1 & NPV1 & 1 & 1 & +0 & 0 & I & -77 573.5\euro & -42 231.2\euro \\
         2 & NPV2 & NPV2 & 1 & 1 & +0 & 0 & I & -113 682.02\euro & -63 680.4\euro\\
         3 & $C_ {tot}$ & $C_{tot}$ & 1 & 1 & +0 & 0 & I & 118 248.66\euro & 66 403.65\euro\\
         4 & CDE & CDE & 1 & 2 & +10 & +1 & D & 18.556 tCO2eq & 68.554 tCO2eq  \\
         5 & ROI & (financial) & 1 & 1 & +0 & 0 & I & 0\% & \\
         6 & ROI & CDE & 1 & 2 & +0 & +1 & C & 0.415\% & 68.786 tCO2eq \\
         7 & (financial) & PkWh* & 1 & $\varnothing$ & / & / & / & / & 0.146\euro/kWh \\
         8 & NPV1 & CDE & 1 & 2 & +0 & +1 & C & -21 553.8\euro & 68.786 tCO2eq \\
         9 & NPV2 & CDE & 1 & 2 & +0 & +1 & C & -29 291.5\euro & 68.786 tCO2eq \\
         10& $C_{tot}$ & CDE & 1 & 2 & +0 & +1 & C & 30 272.4\euro & 68.786 tCO2eq\\
         11& CDE & (financial) & 1 & 1 & +0 & 0 & I & 60 561 tCO2eq &  \\
         12& PkWh* & (financial) & 1 & 1 & +0 & +1 & I & 0.214\euro/kWh & \\
         13& PkWh* & CDE & 1 & 2 & +10 & +1 & D & 0.12\euro/kWh & 68.554 tCO2eq\\
         \hline
    \end{tabular}%}
    \caption{Expected utilities at SPEs of the NMIP $\Gamma^2$ for the REC in surplus with the candidates set $\overline{\newusers}$ and the rationality of agents is assumed. The (financial) notation indicates that the results are valid if one of the 3 criteria: NPV1, NPV2 and $C_{tot}$ is used. We have used the lexicographical order in the case of PkWh.}
    \label{tab:RatSPE-sur-tree2}
    \vspace{-1em}
\end{table}

There is a unique SPE for each case in Tab.\ref{tab:RatSPE-sur-tree2}. In financial case (1-3,5,11-12), the REC includes user 1 without requiring investment (except in 12 with a battery) and increases local prices. Whenever the REC aims to minimize its CDE, it selects user 2 with investment notably in one battery regardless of her criteria. An intriguing outcome occurs in scenario 7, where the REC chooses not to select any candidate. This results stems from a SPE determined under the lexicographic order assumed for the REC, suggesting that alternative SPEs with different outcomes may have been overlooked. Once profile 11 is excluded, both heuristic metrics (Tab.\ref{fig:Metric_Sur}) identify user 1, as the most suitable candidate to join the REC. However, the metrics were unable to predict model's selection of user 2, given that neither metric incorporates CO2 emissions.

%\subsubsection{Observations summary}
%We summarize the observations arising from the comparison between the results of heuristic methods and those of subgames perfect equilibria of the NMIP $\Gamma^2$ (Fig.\ref{fig:Tree2}), where the community moves first.
%\begin{mybox}
%\begin{enumerate}[topsep=2pt,itemsep=0pt,parsep=0pt]
%    \item The characteristics of preferences and their combinations directly influence the outcome of the problem, highlighting the importance of strategic choices for each stakeholder.
%    \item For each preference criterion of the candidates set, if the REC adopts a strictly financial preference (such as NPV1, NPV2 or total cost $C_{tot}$), heuristic methods can help forecast the new member selected by the community.
%    \item When the community pursues objectives that are not purely financial such as minimizing the CDE or the price of kWh, heuristic methods are not always able to anticipate the REC's choice. In such cases, decisions depend on the specific features of the SPEs and on the candidates' preferences.
%\end{enumerate}   
%\end{mybox}

\subsection{Impact of the decisions order}\label{subsec:ResSPE-order-Rat}
We study the impact of decision order on equilibria outcomes. Adopting the perspective of a user $j\in\newusers$ , we compare results from both sequential games $\Gamma^1_j$ (user $j$ moves first, Fig.\ref{fig:Tree1}) and $\Gamma^2$ (REC moves first, Fig.\ref{fig:Tree2}). To limit the scope, we focus on users selected in a SPE of $\Gamma^2$ in Subsection \ref{subsec:heu-vs-spe}, excluding profiles 7 and 11. For brevity, we omit here the analysis of surplus REC , which displays different strategic trade-offs and REC responses. A full discussion is provided in \citet{Sadoine25}.\par
\ppar

Since user 2 was always selected by the energy-deficit REC across all combinations, we compare this user's outcomes in both games. Table \ref{tab:RatSPE-def-tree1.2} shows the actions and expected utilities $U$ \eqref{funct:EUT} from the SPEs of $\Gamma^1_2$.
\renewcommand{\arraystretch}{0.7}
\begin{table}[ht!]
    \centering
    \scriptsize
    %\resizebox{\textwidth}{!}{
    \begin{tabular}{l|ccccccccc}
    \hline
         &\multicolumn{2}{c}{Criteria}& Num. & In & PV & \multirow{2}{*}{Stor.} & Local & Exp. utility & Exp. utility \\
         &User 2 & REC & SPE & REC & (kWp) & & prices & user 2 & REC \\
         \hline
         2' & NPV2 & NPV2 & 1 & 1 & +0 & +1 & D & -29 316.5\euro & -200 652.46\euro\\
         3' & $C_ {tot}$ & $C_{tot}$ & 1 & 1 & +0 & +1 & D & 30 298.4\euro & 208 822\euro\\ \hline
         \multirow{2}{*}{4'} & \multirow{2}{*}{CDE} & \multirow{2}{*}{CDE} & \multirow{2}{*}{2} & 0 & \multirow{2}{*}{+10} & \multirow{2}{*}{+1} & $\backslash$ & \multirow{2}{*}{18.556 tCO2eq} & 137.564 tCO2eq  \\
         & & & & 1 & & & D & & 109.65 tCO2eq\\
         \hline
         \multirow{2}{*}{5'} & \multirow{2}{*}{ROI} & \multirow{2}{*}{(financial)} & \multirow{2}{*}{2} & 0 & \multirow{2}{*}{+0} & \multirow{2}{*}{+1} & $\backslash$ & \multirow{2}{*}{0.409\%} &   \\
         & & & & 1 & & & D & & \\
         \hline
         7' & NPV1 & PkWh* & 1 & 1 & +0 & 0 & I & -19 453.4\euro & 0.219\euro/kWh \\
         8' & NPV2 & PkWh* & 1 & 1 & +0 & +1 & I & -27 456.6\euro & 0.221\euro/kWh \\
         9' & $C_{tot}$ & PkWh* & 1 & 1 & +0 & +1 & I & 28 366.51\euro & 0.221\euro/kWh \\
         10' & NPV1 & CDE & 1 & 1 & +0 & 0 & C & -20 262.8\euro & 123.556 tCO2eq \\
         12' & $C_{tot}$ & CDE & 1 & 1 & +0 & +1 & D & 30 298.4\euro & 126.527 tCO2eq\\
         \hline
         \multirow{2}{*}{13'} & \multirow{2}{*}{CDE} & \multirow{2}{*}{(financial)} & \multirow{2}{*}{2} & 0 & \multirow{2}{*}{+10} & \multirow{2}{*}{+1} & $\backslash$ & \multirow{2}{*}{18.556 tCO2eq} &   \\
          & & & & 1 & & & D & & \\ \hline
         \multirow{2}{*}{14'} & \multirow{2}{*}{PkWh*} & \multirow{2}{*}{(financial)} & \multirow{2}{*}{2} & 0 & \multirow{2}{*}{+10} & \multirow{2}{*}{+1} & $\backslash$ & \multirow{2}{*}{0.12\euro/kWh} &   \\
          & & & & 1 & & & D & & \\
         \hline
    \end{tabular}%}
    \caption{Expected utilities at SPEs of the NMIP $\Gamma^1_2$ for the REC in deficit where the rationality of agents is assumed. The (financial) notation indicates that the results are valid if one of the 3 criteria: NPV1, NPV2 and $C_{tot}$ is used. We have used the lexicographical order in the case of PkWh.}
    \label{tab:RatSPE-def-tree1.2}
    \vspace{-1em}
\end{table}

Since the outcomes are identical in both games for combinations 1, 6, 11 and 15 (Tab.\ref{tab:RatSPE-def-tree2}), they are omitted from Tab.\ref{tab:RatSPE-def-tree1.2}. User 2 seems more inclined to invest in a battery when acting first, as seen in combinations 2', 3', 5', 8', 9' and 12'. For combinations 4', 5', 13' and 14', two SPEs exist, reflecting user 2's indifference between remaining alone and joining the REC.When integration occurs, the actions and outcomes generally match those in $\Gamma^2$, except in case 5'. Across all cases, the REC's expected utility is always lower or equal to that in Tab.\ref{tab:RatSPE-def-tree2}, but still higher than on the baseline scenario with no integration. Thus, although these strategies may not be optimal, they remain beneficial to the REC. These conclusions underline the importance of the decision-making sequential and preference criteria in modeling strategic interactions between RECs and potential new members.

\section{Results with prospect theory}\label{sec:PTvsEUT}
In this section, we examine the LT consequences of decisions made in the NMIP when agents exhibit bounded rationality. This complements the previous analysis (Section \ref{sec:ResRat}) by comparing the outcomes under perfect rationality with those generated through simulations based on prospect theory.\par
\ppar
In order to carry out this evaluation, we consider different parameterizations of stakeholders' PT functions. The first (PT1 in Table \ref{tab:PT-parameters}) follows the widely used estimates from \citet{KT92}: $\eta_a=\eta_b=0.88$, $\eta_c=2.25$ and $\eta_d=0.65$. The second (PT2) adopts the values $\eta_a=\eta_b=0.45$, $\eta_c=1.96$ and $\eta_d=0.65$, as suggested by \citet{Wen10} for organizational decision-making. To our knowledge, no PT parameter calibration exists for RECs; we adopt these two configurations as representative cases for comparison. Reference point choice also significantly influences outcomes. We test two approaches: a fixed reference point $r^{max}$ and a stochastic one $r^{stoc}$ defined in Section \ref{subsec:PT}, to assess their impact on stakeholder decisions. For simplicity, we assume all stakeholders adopt the same reference point type within each scenario. Table \ref{tab:PT-parameters} summarizes the parameter settings across several cases: rational agents, homogeneously and heterogeneously boundedly rationality, and mixed setups with one rational and one non-rational agent. This allows for a nuanced comparison of the effects of behavioral assumptions on strategic outcomes.

\renewcommand{\arraystretch}{0.7}
\begin{table}[ht!]
    \centering
    \scriptsize
    %\resizebox{\textwidth}{!}{
    \begin{tabular}{l|ccccccccccc}
    \hline
    Num. & & & \multicolumn{4}{c}{PT parameters of users} && \multicolumn{4}{c}{PT parameters of REC}\\
    Scenar. & Users $\newusers$& REC & $\eta_a$ & $\eta_b$ & $\eta_c$ & $\eta_d$ && $\eta_a$ & $\eta_b$ & $\eta_c$ & $\eta_d$\\ \hline
    1 & EUT & EUT &      &      &      &      &&      &      &      &     \\
    2 & PT1 & PT1 & 0.88 & 0.88 & 2.25 & 0.65 && 0.88 & 0.88 & 2.25 & 0.65\\
    3 & PT1 & PT2 & 0.88 & 0.88 & 2.25 & 0.65 && 0.45 & 0.45 & 1.96 & 0.65\\
    4 & PT2 & PT1 & 0.45 & 0.45 & 1.96 & 0.65 && 0.88 & 0.88 & 2.25 & 0.65\\
    5 & PT2 & PT2 & 0.45 & 0.45 & 1.96 & 0.65 && 0.45 & 0.45 & 1.96 & 0.65\\
    6 & PT1 & EUT & 0.88 & 0.88 & 2.25 & 0.65 &&      &      &      &     \\
    7 & PT2 & EUT & 0.45 & 0.45 & 1.96 & 0.65 &&      &      &      &     \\
    8 & EUT & PT1 &      &      &      &      && 0.88 & 0.88 & 2.25 & 0.65\\
    9 & EUT & PT2 &      &      &      &      && 0.45 & 0.45 & 1.96 & 0.65\\ \hline
    \end{tabular}%}
    \caption{Scenarios of subjective value and probability weighting functions parameters for users and RECs}
    \label{tab:PT-parameters}
    \vspace{-1.5em}
\end{table}

\subsection{Community new member selection}\label{subsec:ResPT-Tree2}
We compare the SPE outcomes of the extensive-form game $\Gamma^2$ (Fig.\ref{fig:Tree2}) across the parameter configurations in Tab.\ref{tab:PT-parameters}, the two reference point types and for each combination of preference criteria. Thus, we simulate 450 scenarios per REC. Users 7 and 11 are excluded from the candidate list, resulting in a reduced set $\overline{\newusers}$. The following analysis focuses on the most representative preference configurations.

\subsubsection{REC in deficit}
We begin with the REC facing an annual energy deficit. For many preference combinations, applying PT increases the number of resulting SPEs, regardless of the reference point used.
When both stakeholders maximize NPV1 \eqref{funct:NPV} using the fixed reference point $r^{max}$, a non-rational user 2 modeled with PT2 parameters (enterprise level) leads the REC to require an additional PV investment. This increases the REC's expected utility, but does not improve the user's utility. No significant changes occur in the other configurations.\par
\ppar

Table \ref{tab:PTSPE-def-tree2-cas2} shows the actions and outcomes when candidates in $\overline{M}$ minimize their total costs $C_{tot}$ \eqref{funct:Ctot}, while the REC minimizes its CDE $\eqref{funct:CDE}$.
\renewcommand{\arraystretch}{0.7}
\begin{table}[ht!]
    \centering
    \scriptsize
    %\resizebox{\textwidth}{!}{
    \begin{tabular}{l|c|ccccccc}
    \hline
        Ref.& Num. & Num. & \multirow{2}{*}{User} & PV & \multirow{2}{*}{Stor.} & Local & Exp. utility & Exp. utility   \\
        point& Scenar.& SPE & & (kWp) & & prices & user & REC \\ \hline
         $\backslash$ & 1 & 1 & 2 & +0 & 0 & C & 31 410.6\euro & 123.556 tCO2eq \\ \hline \hline
         \multirow{4}{*}{$r^{max}$} & 4,7 & 1 & \multirow{2}{*}{2} & \multirow{2}{*}{+2} & \multirow{2}{*}{0} & \multirow{2}{*}{I} & \multirow{2}{*}{30 341.12\euro} & \multirow{2}{*}{119.912 tCO2eq} \\
         & 5 & 8 & & & & & & \\ \cline{2-9}
         & 2,6,8 & 1 & \multirow{2}{*}{2} & \multirow{2}{*}{+0} & \multirow{2}{*}{0} & \multirow{2}{*}{C} & \multirow{2}{*}{31 410.6\euro} & \multirow{2}{*}{123.556 tCO2eq}\\
         & 3,9 & 8 & & & & & &\\ \hline 
         \multirow{4}{*}{$r^{stoc}$} & 2,4,6,7 & 1 & \multirow{2}{*}{2} & \multirow{2}{*}{+2} & \multirow{2}{*}{0} & \multirow{2}{*}{I} & \multirow{2}{*}{30 341.12\euro} & \multirow{2}{*}{119.912 tCO2eq} \\
         & 3,5 & 32 & & & & & & \\ \cline{2-9}
         & 8 & 1 & \multirow{2}{*}{2} & \multirow{2}{*}{+0} & \multirow{2}{*}{0} & \multirow{2}{*}{C} & \multirow{2}{*}{31 410.6\euro} & \multirow{2}{*}{123.556 tCO2eq}\\
         & 9 & 32 & & & & & & \\ \hline 
    \end{tabular}%}
    \caption{Actions and outcomes at SPEs of game $\Gamma^2$, for candidates minimizing $C_{tot}$ and the energy-deficit REC minimizing its CDE. PT results are shown in terms of expected utility to enable direct comparison.}
    \label{tab:PTSPE-def-tree2-cas2}
    \vspace{-1em}
\end{table}
With fixed reference point $r^{max}$, if user 2 is non-rational with PT2, the REC requires investments in two additional PV panels. This results in higher expected utilities for both stakeholders compared to the rational case. If the REC also follows PT2 (case 5), 8 SPEs emerge with identical actions. A similar deviation occurs with the stochastic reference point, across all case where user 2 follows PT1 or PT2. When both agents follow PT2, 32 SPEs are observed. In other cases, actions and outcomes remains aligned with the rational case, though the number of SPEs may increase.

\subsubsection{REC in surplus}
Except when candidates maximize NPV1 and the energy-surplus REC minimizes CDE, no deviations in actions are observed across the different scenarios and preference combinations. Only minor variations in the number of SPEs occur. The actions and outcomes for this special combination are displayed in Table \ref{tab:PTSPE-sur-tree2-cas1}. With fixed reference points $r^{max}$, if candidates are non-rational (PT1 or PT2), the REC chooses not to include any new member, remaining in its original composition. This outcome benefits user 2, as the investment required in the rational case would reduce her expected utility of NPV1. With stochastic reference point, actions remain unchanged, but the number of SPEs increases to two when the REC follows PT2.
\renewcommand{\arraystretch}{0.7}
\begin{table}[ht!]
    \centering
    \scriptsize
    %\resizebox{\textwidth}{!}{
    \begin{tabular}{l|c|ccccccc}
        \hline
         Ref.& Num. & Num. & \multirow{2}{*}{User} & PV & \multirow{2}{*}{Stor.} & Local & Value & Value   \\
         point& Scenar.& SPE & & (kWp) & & prices & user & REC \\ \hline
         $\backslash$ & 1 & 1 & 2 & +0 & +1 & C & -21 553.8\euro & 68.786 tCO2eq \\ \hline \hline
        \multirow{2}{*}{$r^{max}$}& 2-7 & 1 & $\varnothing$ & $\backslash$ & $\backslash$ & $\backslash$ & $\backslash$ & 69.016 tCO2eq\\
        & 8,9 & 1 & 2 & +0 & +1 & C & -21 553.8\euro & 68.786 tCO2eq \\ \hline
        \multirow{2}{*}{$r^{stoc}$} & 2,4,6,7,8 & 1 & \multirow{2}{*}{2} & \multirow{2}{*}{+0} & \multirow{2}{*}{+1} & \multirow{2}{*}{C} & \multirow{2}{*}{-21 553.8\euro} & \multirow{2}{*}{68.786 tCO2eq}\\
        & 3,5,9& 2 & & & & & &\\
        \hline
    \end{tabular}%}
    \caption{Actions and outcomes at SPEs of game $\Gamma^2$, for candidates maximizing NPV1 and the energy-surplus REC minimizing its CDE. PT results are shown in terms of expected utility to enable direct comparison.}
    \label{tab:PTSPE-sur-tree2-cas1}
    \vspace{-1em}
\end{table}

%\subsubsection{Observations summary}

%We summarize the conclusions of this subsection.
%\begin{mybox}
%In the context of the extensive game $\Gamma^2$ in Fig. \ref{fig:Tree2}:
%\begin{enumerate}[topsep=2pt,itemsep=0pt,parsep=0pt]
%    \item Stakeholders' behavior and decisions may differ between cases of perfect rationality and those subject to bounded rationality via PT.%\tbcom{Referer directement à la prospect theory? "bounded rationnality" peut avoir d'autres sens}
%    \item The choice parameters is a fundamental element of PT to capture the behaviors. It seems crucial to use a parameterization suited to the type of user or entity under study, whether individuals, organizations or energy communities.
%    \item The selection of the reference point is also decisive. 
%    \item The combinations of preferences can lead to differences in actions and results for the community and users.
%\end{enumerate}
%\end{mybox}

%These deviations emphasize the necessity of considering more nuanced behavioral models. Further, they also call for experimental studies with energy communities to estimate suitable parameters for this new mechanism, and thus refine the relevance of behavioral models applied to these specific contexts.

\subsection{Impact of the decisions order with PT framework}
We assess the impact of decision order by comparing results of a user $j\in\overline{M}$ at SPEs from both games $\Gamma^1_j$ and $\Gamma^2$, across various system configurations (Tab.\ref{tab:PT-parameters}) and for each preference combination. In total, 3600 scenarios per REC were simulated, covering all candidates in $\overline{\newusers}$, preference combinations, PT parameterization and both fixed and stochastic reference points. We focus on the most significant users and combinations in the energy-deficit REC. We also examined surplus cases, a discussion is available in \citet{Sadoine25}.\par
\ppar
In Section \ref{subsec:ResPT-Tree2}, the energy-deficit REC still selects user 2 as a new member. Among the tested cases, we focus on the most behaviorally significant one: users minimize total costs \eqref{funct:Ctot}, while the REC minimize its CDE \eqref{funct:CDE}. Table \ref{tab:PTSPE-def-tree1.2} reports the actions and outcomes at SPEs of $\Gamma^1_2$, with the rational benchmark in first row. For comparison, PT outcomes are expressed in terms of expected utility \eqref{funct:EUT}. User 2's rational strategy differs between the two games (Tabs.\ref{tab:PTSPE-def-tree2-cas2} \& \ref{tab:PTSPE-def-tree1.2}). In game $\Gamma^1_2$, when reference points are fixed and user 2 adopts PT2 parameter (organization level), he still chooses to join the REC, even though her expected utility drops while the REC's improves. These results are consistent in $\Gamma^2$ under the same conditions. In other cases, the strategies differ between games. The observations remain similar when reference points are stochastic and user 2's behavior is under PT1 or PT2. In all cases, the REC still benefit from selecting user 2.
\begin{comment}
For the first setting, no change is observed in game $\Gamma^1_2$: all parameterization scenarios and reference points yield the same strategies as in the rational case. Thus, user 2's strategies remain identical in both games $\Gamma^1_2$ and $\Gamma^2$, except the case where the reference point is fixed and user 2 follows the PT2 parameters.
\par
\ppar
\end{comment}

\renewcommand{\arraystretch}{0.7}
\begin{table}[ht!]
    \centering
    \scriptsize
    %\resizebox{\textwidth}{!}{
    \begin{tabular}{l|c|ccccccc}
        \hline
         Ref.& Num. & Num. & In & PV & \multirow{2}{*}{Stor.} & Local & Exp. utility & Exp. utility \\
         point& Scenar. & SPE & REC & (kWp) & & prices & user 2 & REC \\ \hline
         $\backslash$ & 1 & 1 & 1 & +0 & +1 & D & 30 298.4\euro & 126.527 tCO2eq\\ \hline \hline
         \multirow{2}{*}{$r^{max}$} & 4,5,7 & 1 & \multirow{2}{*}{1} & +2 & 0 & I & 30 341.12\euro & 119.912 tCO2eq \\ 
         & 2,3,6,8,9& 1 & & +0 & +1 & D & 30 298.4\euro & 126.527 tCO2eq\\ \hline
         \multirow{4}{*}{$r^{stoc}$} & 2,4,6,7 & 1 & \multirow{2}{*}{1} & \multirow{2}{*}{+2} & \multirow{2}{*}{0} & \multirow{2}{*}{I} & \multirow{2}{*}{30 341.12\euro} & \multirow{2}{*}{119.912 tCO2eq}\\
         & 3,5 & 4 & & & & & & \\
         \cline{2-9}
         & 8 & 1 & \multirow{2}{*}{1} & \multirow{2}{*}{+0} & \multirow{2}{*}{+1} & \multirow{2}{*}{D} & \multirow{2}{*}{30 298.4\euro} & \multirow{2}{*}{126.527 tCO2eq}\\
         & 9 & 4 & & & & & & \\ \hline
    \end{tabular}%}
    \caption{Actions and outcomes at SPEs of game $\Gamma^1_2$, for user 2 minimizing $C_{tot}$ and the energy-deficit REC minimizing its CDE. PT results are shown in terms of expected utility to enable direct comparison.}
    \label{tab:PTSPE-def-tree1.2}
    \vspace{-1em}
\end{table}

 \section{Conclusion}\label{sec:Conclusions}
This paper proposed a novel framework for addressing the relatively unexplored problem of integrating a new member into an existing renewable energy community (REC). The \acrfull{NMIP} models both long-term strategic decisions (e.g., investments, local pricing) and short-term operational decisions (daily energy scheduling) within a DSM-scheme, under exogenous retail price uncertainty. We use extensive game theory to model the different time horizons of the decision-making. In particular, ST decisions of the day-ahead energy scheduling, are modeled using the GNEP studied in \citet{SGB25}. To capture the complex dynamics of the process, two extensive games with distinct decision sequences have been established. One where an external user is interested to join a REC and moves first ($\Gamma^1$, Fig.\ref{fig:Tree1}), and another where the REC initiates its expansion  ($\Gamma^2$, Fig.\ref{fig:Tree2}), and selects a new member among a candidate set $\newusers$. This enabled analysis of how decision order impacts strategies and subgame perfect equilibria. A key strength of the proposed approach lies in its flexibility. The theoretical formulation can be extended to encompass a variety of scenarios and stakeholder preference, ranging from financial (e.g., costs, NPV, ROI, etc.) to environmental objectives (e.g., CO2 emissions). It also supports both perfect and bounded rationality assumptions, including various prospect theory parameterizations and reference point selections. This adaptability opens several avenues for future developments.
\par
\ppar
Simulation results from the case study yield several insights into how preferences criteria, decision order, and behavioral assumptions influence the actions and outcomes of the NMIP. First we compared heuristic methods from \citet{MRDP22a} with the SPEs of game $\Gamma^2$. When the REC adopts financial objectives (e.g., NPV, total cost), heuristic methods can effectively predict the selected candidate by the REC. In contrast, when the REC focuses on carbon emissions or kWh price, they may fail to anticipate the REC's choice, which thus depends on the equilibrium structure and candidate preferences. We observed that the nature of stakeholder preferences, their parameters and combinations impact strategies and outcomes, regardless of the decision sequence or behavioral assumptions. The decision order further influence agents' anticipations, affecting SPEs and stakeholder interests, under both perfect and bounded rationality. Incorporating prospect theory reveals deviations from rational benchmarks and shows that stakeholder behavior is highly sensitive to the choice of reference point and the parameterization used (e.g., individual vs. corporate settings). These effects can lead to counter-intuitive outcomes, such as users opting to join the REC despite lower expected utility, highlighting the importance of accurately modeling subjective perceptions and bounded rationality in strategic decision-making under uncertainty. \par
\ppar
The current models assumes individual investments in the integration framework; a natural extension would consider joint investment decisions or the case of a member leaving the REC under both individual and collective ownership schemes. The actions space could also be expanded beyond discrete profiles to infinite sets, allowing for finer modeling of uncertainties and behaviors. Although our case study used homogeneous preferences and reference point settings, relaxing these assumptions would better reflect real-world heterogeneity in objectives and perceptions.In fact, the proposed framework is flexible enough to handle a wide variety of scenarios, including additional criteria or prospect theory parameterization and reference point selection. From an empirical perspective, validating the relevance of prospect theory in the REC context remains essential to calibrate realist behavioral models. Finally, a key open question concern RECs governance: how should preferences be defined and managed in communities composed of heterogeneous prosumers with divergent goals and risk perceptions? Our approach has so far considered the existing community as a single entity with well-defined collective objectives for LT decision-making. A deeper exploration of this topic could provide interesting perspectives both for theoretical modeling and for the development of adapted regulations.

\bibliographystyle{elsarticle-harv}
\bibliography{Reference}

\section*{CRediT autorship contribution statement}
\textbf{Louise Sadoine:} Writing-original draft, Methodology, Software, Formal analysis, Conceptualization, Visualization. \textbf{Thomas Brihaye:} Writing-review \& editing, Supervision, Methodology, Conceptualization. \textbf{Zacharie De Grève:} Writing-review \& editing, Supervision, Conceptualization, Funding acquisition. 

\section*{Declaration of competing interest}
The authors declare that they have no known competing financial interests or personal relationships that could have been appeared to influence the work reported in this paper.

\section*{Acknowledgments}
This work was supported by the FPS Economy, S.M.E.s , Self- Employed and Energy through the Energy Transition Funds Project ALEXANDER.

\section*{Declaration of generative AI in the manuscript preparation process}
During the preparation of this work the authors used OpenAI's ChatGPT (GPT-5) in order to provide writing assistance, including minor improvements in clarity and style. After using this tool, the authors reviewed and edited the content as needed and takes full responsibility for the content of the final text.

\end{document}

%% file: acronyms.tex
\newacronym{CSC}{CSC}{Collective Self-Consumption}
\newacronym{CDE}{CDE}{Carbon Dioxide Emissions}
\newacronym{DA}{DA}{Day-Ahead}
\newacronym{DER}{DER}{Distributed Energy Resource}
\newacronym{DSM}{DSM}{Demand-Side Management}
\newacronym{ESS}{ESS}{Energy Storage System}
\newacronym{EUT}{EUT}{Expected Utility Theory}
\newacronym{GNE}{GNE}{Generalized Nash Equilibrium}
\newacronym{GNEP}{GNEP}{Generalized Nash Equilibrium Problem}
\newacronym{LT}{LT}{Long-Term}
\newacronym{LV}{LV}{Low Voltage}
\newacronym{MS}{MS}{Matching Score}
\newacronym{NMIP}{NMIP}{New Member Integration Problem}
\newacronym{NPV}{NPV}{Net Present Value}
\newacronym{PT}{PT}{Prospect Theory}
\newacronym{PV}{PV}{Photovoltaic}
\newacronym{REC}{REC}{Renewable Energy Community}
\newacronym{ROI}{ROI}{Return on Investment}
\newacronym{SC}{SC}{Social Cost}
\newacronym{SPE}{SPE}{Subgame Perfect Equilibrium}
\newacronym{ST}{ST}{Short-Term}
\newacronym{VE}{VE}{Variational Equilibrium}